\documentclass{aa}  

\usepackage{graphicx}
\usepackage{txfonts}
\usepackage{url}
\usepackage{multirow}
\usepackage{longtable,lscape}
\usepackage{color}
\usepackage{multirow}
\newcommand{\teff}{$T_{\rm eff}$}
\newcommand{\logg}{$\log g$}
\newcommand{\vsini}{$v\sin i$}

\definecolor{rosso}{rgb}{1,0,0}
\definecolor{blu}{rgb}{0,0,1}

\definecolor{mag}{rgb}{1,0,1}
\definecolor{verde}{rgb}{0,0.6,0}

\begin{document}

\title{Stellar Population Astrophysics (SPA) with the TNG}
\subtitle{Stock\,2, a little-studied open cluster with an eMSTO\thanks{Based on observations made with the Italian Telescopio Nazionale Galileo (TNG) operated on the island of La Palma by the 
Fundaci\'{o}n Galileo Galilei of the INAF (Istituto Nazionale di Astrofisica) at the Observatorio del Roque de los Muchachos. This study is part of 
the Large Program titled SPA -- Stellar Population Astrophysics: the detailed, age-resolved chemistry of the Milky Way disk (PI: L. Origlia), granted 
observing time with HARPS-N and GIANO-B echelle spectrographs at the TNG.}}
\titlerunning{Stock 2}

\author{
J. Alonso-Santiago\inst{1}
\and A. Frasca\inst{1}
\and G. Catanzaro\inst{1}
\and A. Bragaglia\inst{2}
\and G. Andreuzzi\inst{3,4}
\and R. Carrera\inst{5}
\and E. Carretta\inst{2}
\and G. Casali\inst{6,7}
\and V. D'Orazi\inst{5}
\and X. Fu\inst{8}
\and M. Giarrusso\inst{9}
\and S. Lucatello\inst{5}
\and L. Magrini\inst{6}
\and L. Origlia\inst{2}
\and L. Spina\inst{5}
\and A. Vallenari\inst{5}
\and R. Zhang\inst{5,10}
}

\institute{
INAF--Osservatorio Astrofisico di Catania, via S. Sofia 78, 95123 Catania, Italy\\
\email{javier.alonso@inaf.it} \\
\and INAF--Osservatorio di Astrofisica e Scienza dello Spazio, Via P. Gobetti 93/3, 40129 Bologna, Italy\\
\and Fundaci\'{o}n Galileo Galilei--INAF, Rambla Jos\'{e} Ana Fern\'{a}ndez P\'{e}rez 7, 38712 Bre\~{n}a Baja, Tenerife, Spain\\
\and INAF--Osservatorio Astronomico di Roma, Via Frascati 33, 00078 Monte Porzio Catone, Italy\\
\and INAF--Osservatorio Astronomico di Padova, Vicolo dell'Osservatorio 5, 35122 Padova, Italy\\
\and Dipartamento di Fisica e Astronomia, Universit\`{a} degli Studi di Firenze, via G. Sansone 1, 50019 Sesto Fiorentino (Firenze), Italy\\
\and INAF--Osservatorio Astrofisico di Arcetri, Largo E. Fermi 5, 50125 Firenze, Italy\\
\and The Kavli Institute for Astronomy and Astrophysics at Peking University, 100871 Beijing, China\\
\and INFN, Laboratori Nazionali del Sud, Via S. Sofia 62, I-95123 Catania, Italy\\
\and Dipartamento di Fisica e Astronomia, Universit\`{a} di Padova, vicolo Osservatorio 2, 35122 Padova, Italy\\
}

\date{}

% \abstract{}{}{}{}{} 
% 5 {} token are mandatory
 
  \abstract
   {%Open clusters are the natural laboratories in which stellar evolution can be studied. In addition, they allow us to track the chemical evolution and properties 
   %of the Galactic disc. In this context, we focused on 
   Stock\,2 is a little-studied open cluster that shows an extended main-sequence turnoff (eMSTO). 
   In order to investigate this phenomenon and characterise the cluster itself
   we performed high-resolution spectroscopy in the framework of the Stellar Population Astrophysics (SPA) project.
   We employed the High Accuracy Radial velocity Planet Searcher in North hemisphere spectrograph (HARPS-N) at the Telescopio Nazionale Galileo (TNG).
   We completed our observations with additional spectra taken with the Catania Astrophysical Observatory Spectrograph (CAOS). In total we observed 46 stars (dwarfs and giants), which represent, by far, the largest 
   sample collected for this cluster to date. We provide the 
   stellar parameters, extinction, radial and projected rotational velocities for most of the stars. Chemical abundances for 21 species with atomic numbers up to 56 have also 
   been derived. We notice a differential reddening in the cluster field whose average value is 0.27 mag. It seems to be the main responsible for the observed eMSTO, since 
   it cannot be explained as the result of different rotational velocities, as found in other clusters. We estimate an age for Stock\,2 of 450$\pm$150\,Ma which corresponds to 
   a MSTO stellar mass of $\approx$2.8\,M$_{\sun}$. The cluster mean radial 
   velocity is around 8.0\,km\,s$^{-1}$. We find a solar-like metallicity for the cluster, [Fe/H]=$-$0.07$\pm$0.06, 
   compatible with its Galactocentric distance. MS stars and giants show chemical abundances compatible within the errors, with the exceptions of Barium and Strontium, which 
   are clearly overabundant in giants, and Cobalt, which is only marginally overabundant. Finally, Stock\,2 presents a chemical composition fully compatible with that observed in other open clusters of the Galactic thin 
   disc.
   }


   \maketitle
%
%________________________________________________________________

\section{Introduction}\label{intro}

During the last years, a large number of young and intermediate-age stellar clusters (with ages up to around two billion years) have been discovered in the Magellanic
Clouds (MCs) exhibiting extended main-sequence turnoffs \citep[eMSTOs,][]{Mackey07,Milone09,Li17,Milone18}. Among them, the youngest ones ($\tau$\,$\leq$\,700\,Ma)
also display split main sequences \citep[MSs,][]{Bastian17,Correnti17,Li17,Milone18}, similar to those observed in the old globular clusters of the Milky Way (MW). 
These features are not a peculiarity only of the MCs clusters but they have recently been found in Galactic open clusters as well \citep{Marino18a,Cordoni18,Piatti19,Li19,Sun19}.
This fact, which appears to be quite common, leads us to critically reconsider the assumption that colour-magnitude diagrams (CMDs) of open clusters can be reproduced
by a single isochrone, as a consequence of an unique and homogeneous stellar population, as it was thought until now. This has led, for instance,
to the use of the so-called isochrone cloud to fit the CMDs of cluster displaying eMSTOs \citep{Johnston19}.

It has been observed that the magnitude of the eMSTO/split MS phenomenon is related to the cluster age \citep{Niederhofer15,Cordoni18}, which would imply that behind
it exists an evolutionary effect. Stellar rotation is accepted as the main responsible \citep{Marino18b,Sun19}. By comparing observed and synthetic CMDs, split MSs have been explained by the 
coexistence of two stellar populations with different rotation rates \citep{D'Antona15,Milone16}. One of them, which includes around two-thirds of the total MS stars, consists of fast rotators 
and forms the so-called red MS (rMS), while the other one, the blue MS (bMS), is composed of the slow-rotating stars. Additionally, in the area of the CMDs around the MSTO, fast rotators 
%occupy higher positions since they 
are brighter than the slow ones. This picture has been confirmed directly from the measurement of projected rotational velocities ($v$\,sin\,i) among eMSTO stars
in both MCs \citep{Dupree17, Marino18b} and MW open clusters \citep{Sun19}.

However, the rotation alone is not always able to explain the observational behaviour and in certain situations 
an age spread, resulting from a prolonged star formation history or multiple star formation episodes,
is also required \citep{Goudfrooij17,Gossage19}. Nonetheless, this is not the case for open clusters, whose mass is well below that 
considered necessary to originate multiple populations \citep{Krumholz2019,Gratton19}.
Alternatively, according to \citet{D'Antona17} the rotational braking due to tidal interactions between the components of close binaries
from a single stellar population of coeval stars, may also produce a distribution of rotational velocities capable to reproduce the eMSTOs and split MSs observed in the CMDs.   
A greater number of observations are necessary to elucidate and constrain the role of each of these mechanisms, or any other that is still hidden underneath, that allows us
to fully understand this phenomenon.

Here we report the analysis of a large sample of stars, both on the MS and giants in the nearby and poorly studied open cluster Stock\,2.
It is a dispersed cluster discovered by \citet{Stock56} located in the Orion spiral arm, [$\alpha$(2000)\,=\,2h15m, 
$\delta$(2000)\,=\,+59$^{\circ}$16$'$, $\ell$\,=\,133.334$^{\circ}$, $b$\,=\,-1.694$^{\circ}$\footnote{nominal coordinates according to the WEBDA database, 
\url{https://webda.physics.muni.cz/}}], 
roughly in the same line of sight as the double cluster $h$ $\&$ $\chi$ 
Persei, but considerably closer to the Sun. However, despite its proximity, physical parameters for this cluster such as age or chemical composition are not precisely known. 
According to the literature \citep{Stock56,Krzeminski1967,Robichon1999,Spagna09} the distance to Stock\,2 ranges between 300 and 350 pc, although the most recent 
studies, based on the second $Gaia$ data release, place it
%slightly below
at about 400 pc \citep{Cantat2018,Reddy2019}. The average reddening is $E(B-V)$\,$\approx$\,0.35, but it seems to
be variable across the cluster field \citep{Krzeminski1967,Spagna09,Ye21}.

Regarding the age, it is still not precisely known. On the one hand, the cluster might be coeval or slightly older than the Pleiades \citep[100--275 Ma, e.g.][]{Krzeminski1967,Robichon1999,Reddy2019,Ye21} 
but on the other hand, \citet{Sciortino2000}, from the analysis of the cluster X-ray luminosity function, found it to have and age similar to the Hyades ($\tau\simeq$\,625 Ma). \citet{Spagna09}, based on 
the TO region shape and the distribution of the giants on the CMD, reported an age within the 200--500 Ma range. Thus, the age of Stock\,2 is still a debated
issue and represents a challenging task.
Recently, \citet{Reddy2019} performed the first detailed spectroscopic analysis of this cluster so far. They took high-resolution spectra of three red giants, from which they estimated 
a solar-like mean metallicity ([Fe/H]=$-$0.06$\pm0.03$) and the chemical abundances for 23 elements.
\citet{Ye21} obtained a similar value ([Fe/H]=$-$0.04$\pm0.15$) from LAMOST mid-resolution spectra of almost 300 likely members. They also found that Stock\,2 is a massive 
cluster ($\approx$\,4000\,M$_{\sun}$).

The present paper is part of the Stellar Population Astrophysics (SPA) project, an ongoing Large Programme running on the 3.6-m Telescopio Nazionale Galileo (TNG) at
the Roque de los Muchachos Observatory (La Palma, Spain). SPA is an ambitious project whose aim is to reveal the star formation and chemical enrichment history of
the Galaxy, obtaining an age-resolved chemical map of the solar neighbourhood and the Galactic thin disc. More than 500 nearby, representative stars are being observed 
at high resolution in the optical and near-infrared bands by combining the High Accuracy Radial velocity Planet Searcher in North hemisphere spectrograph (HARPS-N) and GIANO-B spectrographs 
\citep[see][for more details on SPA]{Origlia19}.
In this work, we combine high-resolution spectroscopy, archival photometry and the $Gaia$ early third data release \citep[$Gaia$-eDR3,][]{eDR3} in order to investigate the properties 
of Stock\,2, paying special attention to the upper MS and MSTO. The analysis of stellar parameters, CMDs, and  the Lithium abundance are of great importance to 
%help reveal 
constrain the cluster age. The paper is structured as follows. 
In Sect.~\ref{sec_targets} we present our observations and explain the criterium followed to select our targets. Then, in Sect.~\ref{sec_spec_anal} we describe our
spectral analysis and display the results derived: radial velocities, atmospheric parameters and chemical abundances. The determination of the extinction and the 
analysis of the CMDs are detailed in Sect.~\ref{sec_redd} and Sect.~\ref{sec_CMD}, respectively. The discussion and comparison of our results with the literature are conducted in
Sect.~\ref{sec_discuss}. Finally, we summarise our results and present our conclusions in Sect.~\ref{sec_concl}.

%%%%%%%%%%%%%%%%%%%%%%%%%%%%%%%%%%%%%%%%%%%%%%%%%%%%%%%%%%%%%%%%%%%%%%%%%%%%%%%%%%%%%%%%%%%%%%%%%%%%%%%%%%%%%%%%%%%%%%%%%%%%%%%%%%%%%%%%%%%%%%%%%%%%%%%%%%%

\section{Observations and targets selection}\label{sec_targets}

\begin{figure} 
  \centering         

  \includegraphics[width=\columnwidth]{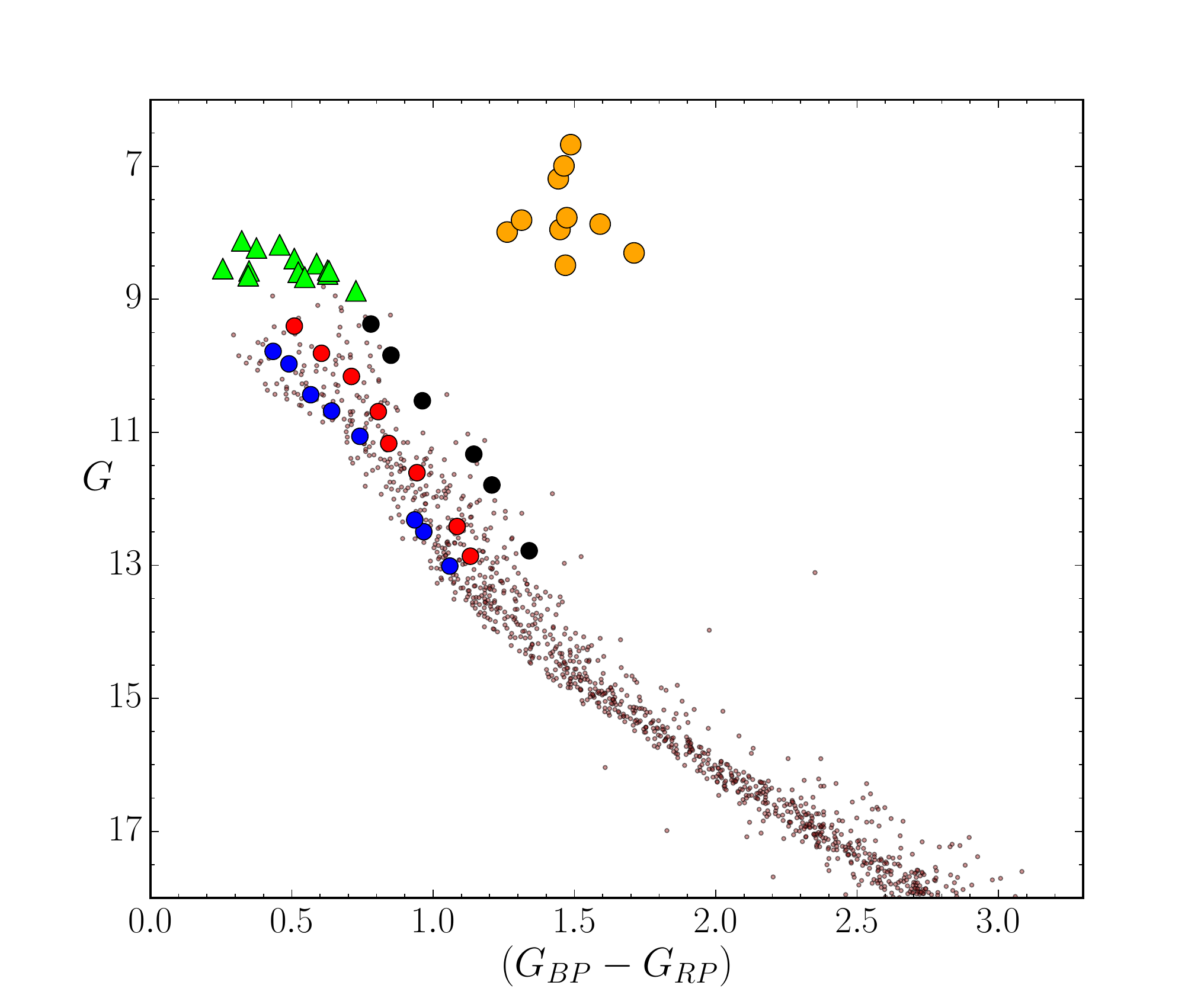}  
  \caption{$G/(G_{\textrm{BP}}-G_{\textrm{RP}})$ diagram for Stock\,2. Members from \citet{Cantat2018} are marked with light brown dots. 
  Stars observed with CAOS in this work are represented with green triangles while those observed with HARPS-N appear as circles with different
  colours, as explained in the text.} 
  \label{fig_targets} 
\end{figure}

With the aim of studying the cluster and determine its properties, we observed a sample of representative stars among the bona-fide members (with an
assigned membership probability of $P$=1) from \citet{Cantat2018}. The only exception is the brightest giant, star g1, for which \citet{Cantat2018} report a
membership probability of $P$=0.8.
We targeted initially the giants, to determine the cluster metallicity and 
detailed abundances, as we did for other clusters in SPA, for which we mainly selected red clump stars, to have a sample as homogeneous as possible 
\citep[see][]{Casali20,Zhang21}. These stars, orange circles in Fig.~\ref{fig_targets}, are labelled as `g' in Table~\ref{tab_obs}. By examining the $Gaia$-DR2 CMD
(since the $Gaia$-eDR3 was not available when we prepared our observations) we realised that the cluster exhibited an eMSTO/split MS, something that was not clearly 
visible in pre-existing photometry, due to field contamination. In order to study it we selected as targets also the brightest stars in the upper MS, close to the turn-off (TO) 
point (green triangles in Fig.~\ref{fig_targets} and labelled as `to' in Table~\ref{tab_obs}) as well as MS stars following three different sequences to sample the blue MS (bMS, 
blue circles and `b'), red MS (rMS, red circles and `r') and the upper envelope of the main sequence, which is the region mostly populated by binary and multiple stars 
(black circles and `u'). The numbering used throughout this paper consists, for each of these series, of assigning a sequential number beginning with the brightest star. 
In total, we acquired high-resolution spectra for 46 stars in several observational runs which are described below (see Table~\ref{tab_obs}).

\subsection{Spectroscopy}

We used HARPS-N \citep{HARPS-N} to observe the ten cluster giants on November 5 and 6, 2018. HARPS-N is an \'{e}chelle spectrograph mounted at the 3.6-m TNG telescope at El Roque de los Muchachos Observatory 
(La Palma, Spain). It is fibre-fed from the Nasmyth B focus and covers the wavelength range from 3870\,\AA{} to 6910\,\AA{} providing a resolving power of 
$R=115\,000$. Later, still with the same equipment, we took spectra for 24 MS stars from 16 to 19 December 2018 and from 13 to 15 January 2019\footnote{
We used GIARPS, i.e. the combination of GIANO and HARPS-N; however, we use only HARPS-N spectra here, as they are more efficient for the warm, MS stars. GIANO spectra 
will be used in forecoming papers}. The instrument's pipeline was used to reduce these spectra.

We completed the TNG observations by collecting additional spectra for the 14 brightest stars of the upper MS around the TO point. Observations were
carried out between 29 and 31 October 2020 with the Catania Astropysical Observatory Spectrograph \citep[CAOS,][]{CAOS1,CAOS2}. 
CAOS is an \'{e}chelle spectrograph mounted on the 0.91-m telescope at {\it M. G. Fracastoro station} (Serra La Nave, Mt Etna (Italy)) which provides a resolution of $R=55\,000$. It is fibre-fed from the
Cassegrain focus and covers, in 81 orders, the wavelength range from 3875\,\AA{} to 6910\,\AA{}. These spectra were reduced by employing the {\scshape iraf}\footnote{{\scshape iraf}
is distributed by the National Optical Astronomy Observatories, which are operated by the Association of Universities for Research in Astronomy, Inc., under the cooperative agreement
with the National Science Foundation.} packages following standard procedures. 
The log of the observations can be found in Table~\ref{tab_obs}. This table displays the spectrograph used, the heliocentric Julian day at mid exposure (HJD), the exposure time ($t_{\textrm{exp}}$, which is the sum of all exposures of the same star),
an estimate of the average signal-to-noise ratio per pixel achieved at 6500\,\AA\ ($S/N$) and the HD (or Tycho, or 2MASS) designation (Name).

\begin{table}[thb!]
\caption{Observation log. }

\begin{center}
\begin{tabular}{lcccc}  
\hline\hline
Star &   Name  & HJD & $t_{\textrm{exp}}$ (s) & $S/N^{a}$  \\
\hline
\noalign{\smallskip}
\multicolumn{5}{c}{\bf HARPS-N}\\
\noalign{\smallskip}
\hline
b1  &    HD\,13967         & 58469.459 & 3000  & 111 \\ 
b2  &    HD\,13100         & 58469.420 & 3000  & 99 \\ 
b3  &    TYC\,3698-2381-1  & 58471.426 & 3600  & 98 \\ 
b4  &    TYC\,3699-1132-1  & 58472.519 & 7200  & 90 \\
b5  &    TYC\,3698-2224-1  & 58499.390 & 3800  & 82 \\ 
b6  &    TYC\,3698-483-1   & 58498.383 & 5400  &  58 \\ 
b7  &   {\tiny J02192173+5927303$^{b}$}  & 58470.375 & 9600  & 64 \\
b8  &   {\tiny J02204032+5923204$^{b}$}  & 58497.360 & 5400  &  30 \\
    &                      &           &       &        \\
r1  &    HD\,12920         & 58499.473 & 1900  &  99 \\
r2  &    TYC\,3698-861-1   & 58469.498 & 3000  &  93 \\ 
r3  &    TYC\,3698-645-1   & 58471.506 & 3600  & 103 \\ 
r4  &    TYC\,3698-2739-1  & 58471.644 & 4800  &  67  \\ 
r5  &    TYC\,3697-479-1   & 58499.438 & 3800  &  78 \\ 
r6  &    {\tiny J02134650+5923569$^{b}$}  & 58498.450 & 5400  &  74 \\ 
r7  &    TYC\,3697-1499-1  & 58470.514 & 9600  &  61 \\
r8  &    {\tiny J02131100+5945191$^{b}$} & 59178.387 & 6300  & 46  \\
    &                      &           &       &   \\
u1  &    HD\,13699         & 58469.381 & 2400  & 147 \\  
u2  &    TYC\,3698-1363-1  & 58469.537 &  3000  & 108 \\ 
u3  &    TYC\,3698-1420-1  & 58471.562 & 4800  & 95 \\ 
u4  &    TYC\,3698-1703-1  & 59131.598 & 3680 &  65   \\
u5  &    {\tiny J02134467+5933039$^{b}$} & 59131.687 & 5520 &  73   \\
u6  &    {\tiny J02162746+5954309$^{b}$} & 59131.748 & 4200 &  28   \\
    &                      &           &       &       \\
g1  &    HD\,15498         & 58428.407 & 700  & 264 \\
g2  &    HD\,14346         & 58428.390 & 700  &  232 \\  
g3  &    HD\,13437         & 58428.468 & 1400  & 346 \\  
g4  &    HD\,13207         &  58428.450 & 1400  & 255 \\  
g5  &    HD\,14403         &  58428.487 & 1400  & 282 \\ 
g6  &    HD\,12650         &  58428.423 & 1400  & 248 \\
g7  &    HD\,15665         &  58429.341 & 1400  & 242 \\
g8  &    HD\,14415         &  58428.505 & 1400  & 255 \\ 
g9  &    HD\,13655         &  58429.359 & 1400  & 211 \\ 
g10 &    HD\,13134         &  58429.378 & 1400  & 192 \\ 
\hline
\noalign{\smallskip}
\multicolumn{5}{c}{\bf CAOS}\\
\noalign{\smallskip}
\hline
to1  &   HD\,14183     &  59152.498  & 2400  & 164 \\
to2  &   HD\,14161     &  59152.567  & 2700  & 153 \\ 
to3  &   HD\,12184     &  59152.529  & 2700  & 164 \\ 
to4  &   HD\,14025     &  59152.601  & 2700  &  89 \\ 
to5  &   HD\,13518     &  59153.510  & 2400  & 114 \\ 
to6  &   HD\,15240     &  59152.474  & 3000  &  70 \\ 
to7  &   HD\,13591     &  59153.541  & 2700  &  97 \\ 
to8  &   HD\,14946     &  59154.361  & 3000  &  69 \\ 
to9  &   HD\,14579     &  59153.615  & 2700  &  81 \\ 
to10 &   HD\,13909     &  59154.489  & 3000  & 153 \\ 
to11 &   HD\,13688     &  59153.576  & 2700  &  80 \\ 
to12 &   HD\,15315     &  59154.404  & 3000  & 115 \\ 
to13 &   HD\,13899     &  59154.526  & 3000  & 153 \\ 
to14 &   HD\,13606     &  59154.323  & 3000  &  45 \\ 
\hline
\end{tabular}
\end{center}
{\bf Notes.} $^{a}$ Signal-to-noise ratio per pixel at 6500\,\AA. $^{b}$ 2MASS designation.
\label{tab_obs}
\end{table}

\subsection{Archival data}

\begin{figure} 
  \centering         

  \includegraphics[width=\columnwidth]{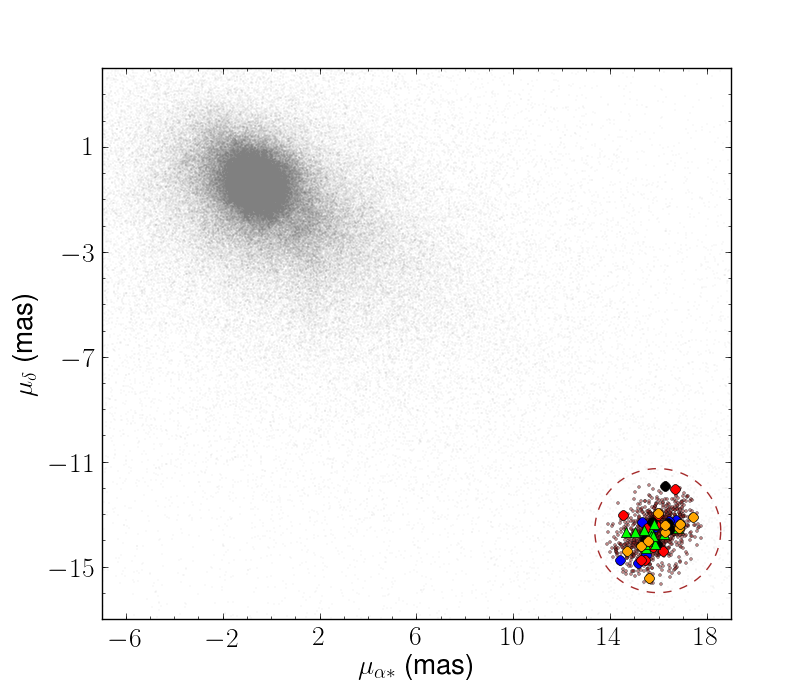}  
  \caption{Proper-motion diagram in the field of Stock\,2. 
  The ellipse (brown dashed line) is centred in the average proper motions of the cluster and has semi axes of 4 times the sigmas of the $\mu_{\alpha*}$ and $\mu_{\delta}$ 
  distributions of the cluster members according to \citet{Cantat2018}. It represents the cluster extent in the astrometric space. 
  Grey dots are field stars whereas the rest of the symbols are the same as those in Fig.~\ref{fig_targets}.} 
  \label{fig_mp}  
\end{figure}

\begin{figure} 
  \centering         

  \includegraphics[width=\columnwidth]{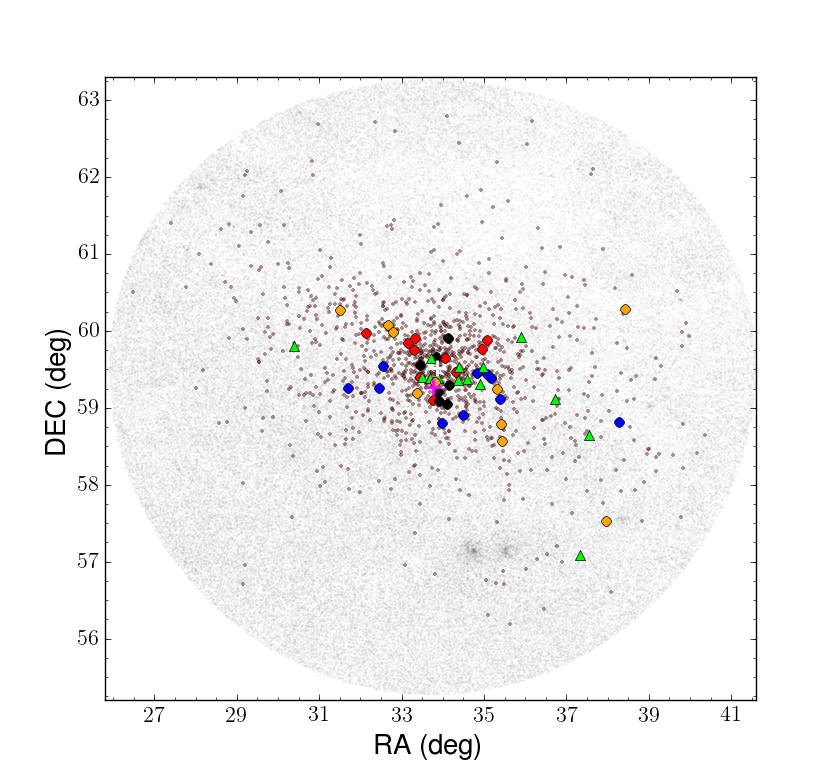}  
  \caption{Sky region around Stock\,2. Grey dots are the sources with $G$\,$\leq$\,16\,mag within a radius of 240\,$\arcmin$ around the cluster nominal 
  centre (magenta cross). Cluster members identified by \citet{Cantat2018} are represented by black points whereas the cluster centre derived from them is
  the white cross. Coloured circles and green triangles are the objects observed in this work (see Fig.~\ref{fig_targets}) with the HARPS-N and CAOS 
  spectrographs, respectively. The overdensities visible at RA$\sim35^{\circ}$ and DEC$\sim57^{\circ}$ correspond to $h$ \& $\chi$ Per double cluster.} 
  \label{fig_distribution} 
\end{figure}

As mentioned above we started our investigation based on the work conducted by \citet{Cantat2018}. From the analysis of $Gaia$-DR2 data they identified 1209 members 
for Stock\,2. In the astrometric space, they located the cluster at ($\mu_{\alpha*}$, $\mu_{\delta}$, $\varpi$) = (15.966, $-$13.627, 2.641) $\pm$ (0.650, 0.591, 0.076),
clearly standing out from the background (as seen in Fig.~\ref{fig_mp}, which highlights the stars observed in this work).
According to the spatial distribution of its members (Fig.~\ref{fig_distribution}) \citet{Cantat2018} placed the cluster centre at $\alpha$(2000)\,=\,2h15m25.44s, 
$\delta$(2000)\,=\,+59$^{\circ}$31$'$19.2$''$, at a distance $\Delta(\alpha,\delta)$=(25.4$^s$,15.3$^{\arcmin}$) from the nominal value. Stock\,2 is a dispersed cluster
and half of its members ($r_{\textrm{50}}$) are found within a radius of 1.03$^{\circ}$ around the centre, with the most distant ones positioned almost 4$^{\circ}$ away. 
As a result, none of the photometric datasets existing in the literature cover its entire extension. For this reason, to complement our spectroscopy and the $Gaia$ data 
we resorted to all-sky photometric surveys. We used $JHK_{\textrm{S}}$ magnitudes from the 2MASS catalogue in the near infrared wavelength \citep{2MASS}
as well as $BVg'r'i'$ optical bands from the APASS catalogue \citep{apass}. In some cases, for the brightest stars for which the APASS photometry is not reliable we
also made use of the values listed in the ASCC2.5 catalogue \citep{ASCC25}.
The combination of all these data allowed us to analyse the CMDs of the cluster, as will be explained later in Sect.~\ref{sec_CMD}. 
All the astrometric and photometric data available for the stars observed in this work are summarised in Tables~\ref{tab_mp} and \ref{tab_fotom} in the appendix of the paper.

%%%%%%%%%%%%%%%%%%%%%%%%%%%%%%%%%%%%%%%%%%%%%%%%%%%%%%%%%%%%%%%%%%%%%%%%%%%%%%%%%%%%%%%%%%%%%%%%%%%%%%%%%%%%%%%%%%%%%%%%%%%%%%%%%%%%%%%%%%%%%%%%%%%%%%%    

\section{Spectral analysis}\label{sec_spec_anal}

\subsection{Radial velocity}

We started the spectroscopic analysis by measuring the heliocentric radial velocity (RV) of the observed objects. For this purpose we cross-correlated our spectra 
against synthetic templates by employing the task {\scshape fxcor} contained in the {\scshape iraf} packages. 
When examining the cross-correlation function (CCF) we identified some multiple systems (SB2 or SB3) among the stars forming our sample namely, r4, u1 and u2. Therefore,
in the upper sequence, we found only two binaries out of the six candidates, although the remaining four could be single-lined systems (SB1).
Additionally, star u3 might also have a close companion since it shows a discrepant {\scshape RUWE\footnote{\url{https://www.cosmos.esa.int/web/gaia/II-124}}} $Gaia$ parameter for a single
source ($\approx$\,3.3).
For the remaining single stars results are listed in the last column of Table~\ref{tab_params}. As can be seen, RVs show a large dispersion, with values ranging from $-$16.5 
to $+$15.7\,km\,s$^{-1}$. This is likely a consequence of the 
$v$\,sin\,$i$ distribution. Indeed, while for slow rotators (e.g. giants and stars in the lower main-sequence) is possible determine
precise RVs, for rapid rotators, instead, it is not. This is specially relevant for the hottest stars in our sample, located at the upper MS close to the TO point. 
These stars, with spectral types A, in addition to rotating rapidly, display far fewer features in their spectra, which broaden and reduce the intensity of the CCF peak.
For this reason, to calculate the average RV for the cluster we only took the stars whose $v$\,sin\,$i$<50\,km\,s$^{-1}$ into account. In this way, from 21 members, we derived 
an average value of $RV$=7.5$\pm$3.3\,km\,s$^{-1}$. On the other hand, $Gaia$-DR2 (since the eDR3 does not provided new values) gives RV for 194 objects among the members listed in \citet{Cantat2018}. 
The average value, after applying a 3$\sigma$-clipping filter to ignore outliers, is $RV_{\textrm{GDR2}}$=9.5$\pm$3.3\,km\,s$^{-1}$ (which becomes 8.0\,km\,s$^{-1}$ if, instead, 
the error-weighted mean is calculated). If we consider only the giants, the weighted average of our values is $RV$=7.9$\pm$1.4\,km\,s$^{-1}$ (where we have assumed the weighted standard deviation 
as uncertainty), which is in close agreement with the above estimate.

\subsection{Atmospheric parameters}\label{sec_params}

To determine the stellar atmospheric parameters of our targets we used the {\scshape rotfit} code \citep{Rotfit} adapted to the SPA project workframe, as 
previously done \citep[see e.g.][]{ASCC123, Casali20}. The code provides us not only with atmospheric parameters such as effective temperature ($T_{\textrm{eff}}$), 
surface gravity ($\log\,g$) and iron abundance ([Fe/H], as a proxy of the metallicity) but also with an estimate of the spectral type (SpT) and the projected rotational
velocity ($v\,\sin\,i$). It should be noted that the last is a key parameter for the research we are conducting in this work.
{\scshape rotfit} is based on a $\chi^2$ minimization of the difference between the target spectrum and a grid of templates. 
This difference is evaluated in 28 spectral segments of 100\,\AA{} each. Then, the final parameters are obtained by averaging the results of the individual regions, weighting them
according to the $\chi^2$ and the information contained in each spectral segment. As template spectra we selected a collection of high-resolution spectra of real stars 
with well-known parameters taken with ELODIE ($R$\,=\,42\,000). This grid of templates is the same as that used in the $Gaia$-ESO Survey by the Catania node
\citep{Smiljanic14, Frasca15}. A more detailed description of our methodology can be found in \citet{ASCC123}.

For all the single stars, the results are displayed in Table~\ref{tab_params}.
We obtained for this cluster an average solar metallicity of [Fe/H]=0.00$\pm$0.08, which was calculated as the weighted mean of the values for the spectra analyzed with {\scshape rotfit}. The 
error reflects the standard deviation of the individual values around the cluster mean.

{\scshape rotfit} is optimised to be used with FGK-type targets. Therefore, for hotter stars we used a different approach based on a grid of synthetic spectra computed as described in 
Sect.~\ref{sec_abund}, for which we adopted an Opacity Distribution Function (ODF) computed for solar abundances.  To determine \teff\ and \logg\ we used the wings and the cores of Balmer lines, 
while a region around the \ion{Mg}{ii}$\lambda$4481 line has been used to  derive the \vsini. Due to the rapid stellar rotation, spectral lines are very broadened and shallow and then very difficult 
to measure, thus we have chosen to adopt [Fe/H]\,=\,0.

\begin{table*}[ht!]
\caption{Stellar parameters derived for the single stars.}
\label{tab_params}
\begin{center}
\begin{tabular}{lrcrlrr} 
\hline\hline
Star  &  $T_{\textrm{eff}}$ (K)~~~~ & $\log\,g$ & [Fe/H]~~~~ & \ Sp T  &  $v \sin\,i$ (km\,s$^{-1}$) & RV (km\,s$^{-1}$) \\  
\hline
b1   &  8500 $\pm$ 300  &  4.10 $\pm$ 0.20  &    0.00$^{a}$~~~~  &  A1\,V$^{b}$  & 140 $\pm$ ~15 &   13.44 $\pm$ 2.69  \\
b2   &  8700 $\pm$ 200  &  4.00 $\pm$ 0.20  &    0.00$^{a}$~~~~  &  A0\,V$^{b}$  &  34.3 $\pm$ 4.7  &    9.37 $\pm$ 0.68  \\
b3   &  7700 $\pm$ 300  &  4.07 $\pm$ 0.23  & $-$0.19 $\pm$ 0.18    &  A7\,V     & 280 $\pm$ ~30 &   15.65 $\pm$ 9.16  \\
b4   &  7800 $\pm$ 300  &  4.09 $\pm$ 0.21  & $-$0.21 $\pm$ 0.16    &  A7\,V            & 120 $\pm$ ~15 &    8.03 $\pm$ 2.85  \\
b5   &  7289 $\pm$ 252  &  4.05 $\pm$ 0.22  & $-$0.16 $\pm$ 0.13    &  A9\,IV           & 220 $\pm$ ~20 &    6.43 $\pm$ 8.04  \\
b6   &  6132 $\pm$ ~~91   &  4.11 $\pm$ 0.14  &    0.00 $\pm$ 0.10    &  F9\,IV-V       &  21.9 $\pm$ 0.7  &    8.47 $\pm$ 0.23  \\
b7   &  6092 $\pm$ ~~73   &  4.20 $\pm$ 0.10  &    0.07 $\pm$ 0.10    &  F8\,V          &   4.0 $\pm$ 1.0  & $-$4.27 $\pm$ 0.10  \\
b8   &  5841 $\pm$ ~~86   &  4.42 $\pm$ 0.12  &    0.06 $\pm$ 0.08    &  G1\,V          &   2.2 $\pm$ 1.7  &    5.74 $\pm$ 0.11  \\
     &                  &                   &                       &                 &                  &                     \\
r1   &  8000 $\pm$ 250  &  3.90 $\pm$ 0.20  &    0.00$^{a}$~~~~  &   A1\,V$^b$     & 250 $\pm$ 30   & $-$1.42 $\pm$ 6.35  \\
r2   &  8300 $\pm$ 300  &  3.80 $\pm$ 0.30  &    0.00$^{a}$~~~~  &  A1\,V$^b$      & 230 $\pm$ 30   &    0.62 $\pm$ 3.09  \\
r3   &  8800 $\pm$ 300  &  3.90 $\pm$ 0.20  &    0.00$^{a}$~~~~  &  A0$^b$         &  40 $\pm$ ~~9  & $-$3.69 $\pm$ 0.54  \\
r5   &  7607 $\pm$ 279  &  4.11 $\pm$ 0.20  & $-$0.09 $\pm$ 0.12    &   F0\,III       & 135 $\pm$ 15 &   13.75 $\pm$ 5.81  \\
r6   &  6851 $\pm$ 138  &  4.14 $\pm$ 0.11  & $-$0.07 $\pm$ 0.09    &   F4\,V         &  13.3 $\pm$1.0  &    9.65 $\pm$ 0.24  \\
r7   &  6332 $\pm$ 163  &  4.04 $\pm$ 0.15  & $-$0.06 $\pm$ 0.11    &  F7\,IV         &  42 $\pm$ ~~2  &    3.55 $\pm$ 0.83  \\
r8   &  6086 $\pm$ ~~73   &  4.20 $\pm$ 0.10  &    0.07 $\pm$ 0.10    &  F8\,V          &   8.5 $\pm$0.8  &    9.02 $\pm$ 0.14  \\                  \\
     &                  &                   &                       &                 &                  &                     \\
u3   &  7603 $\pm$ 299  &  4.01 $\pm$ 0.21  & $-$0.13 $\pm$ 0.12    &  A8\,V         & 53 $\pm$ ~~6   &    9.26 $\pm$ 0.86  \\
u4   &  8300 $\pm$ 300  &  4.10 $\pm$ 0.20  &    0.00$^{a}$~~~~  &  B8\,V$^b$      & 85 $\pm$ 10  & $-$4.96 $\pm$ 0.29  \\
u5   &  6449 $\pm$ 152  &  4.09 $\pm$ 0.16  & $-$0.07 $\pm$ 0.11    &  F6\,IV         & 44 $\pm$ ~~1  &    3.38 $\pm$ 0.75  \\
u6   &  6534 $\pm$ 131  &  4.11 $\pm$ 0.15  & $-$0.05 $\pm$ 0.10    &  F8\,V          & 41 $\pm$ ~~1   &    2.27 $\pm$ 0.76  \\
     &                  &                   &                       &                 &                  &                     \\
g1   &  4530 $\pm$ ~~86   &  2.14 $\pm$ 0.10  &    0.01 $\pm$ 0.09    &   K1\,III       &  1.6 $\pm$ 1.5   &    9.78 $\pm$ 0.12  \\  
g2   &  4760 $\pm$ 111  &  2.69 $\pm$ 0.14  &    0.02 $\pm$ 0.10    &   K0\,III       &  7.6 $\pm$ 0.6   &    8.11 $\pm$ 0.13  \\
g3   &  4937 $\pm$ 114  &  2.51 $\pm$ 0.35  &    0.04 $\pm$ 0.08    &   G8\,III       &  6.1 $\pm$ 0.7   &    8.36 $\pm$ 0.13  \\
g4   &  4977 $\pm$ 117  &  2.82 $\pm$ 0.18  &    0.04 $\pm$ 0.08    &   G8\,III       &  1.7 $\pm$ 1.5   &    8.37 $\pm$ 0.11  \\
g5   &  5061 $\pm$ ~~56   &  2.99 $\pm$ 0.19  &    0.04 $\pm$ 0.07    &   G8\,III       &  5.4 $\pm$ 1.2   &    9.20 $\pm$ 0.12  \\
g6   &  5002 $\pm$ 110  &  2.96 $\pm$ 0.20  &    0.03 $\pm$ 0.07    &   G8\,III       &  1.9 $\pm$ 1.6   &    7.10 $\pm$ 0.10  \\
g7   &  5058 $\pm$ ~~56   &  2.97 $\pm$ 0.20  &    0.03 $\pm$ 0.07    &   G8\,III       &  2.7 $\pm$ 1.6   &    8.45 $\pm$ 0.11  \\
g8   &  5065 $\pm$ ~~56   &  3.00 $\pm$ 0.19  & $-$0.03 $\pm$ 0.09    &   G8\,III       &  5.2 $\pm$ 1.3   &    7.86 $\pm$ 0.11  \\
g9   &  5062 $\pm$ ~~56   &  3.00 $\pm$ 0.19  &    0.00 $\pm$ 0.09    &   G8\,III       &  4.6 $\pm$ 1.4   &    4.38 $\pm$ 0.11  \\
g10  &  5066 $\pm$ ~~56   &  3.01 $\pm$ 0.19  & $-$0.03 $\pm$ 0.09    &   G8\,III       &  4.5 $\pm$ 0.9   &    8.66 $\pm$ 0.11  \\
     &                  &                   &                       &                 &                  &                     \\
to1$^c$  &  9300 $\pm$ 300 &  4.5 $\pm$ 0.2 &   0.00$^{a}$~~~~  &   A1\,V$^{b}$   &   80 $\pm$ 10    &     4.0 $\pm$ 11.7  \\ % SpT de SIMBAD
to2  &  9100 $\pm$ 300  &  4.3 $\pm$ 0.2    &   0.00$^{a}$~~~~  &   A2\,IV$^{b}$  &   60 $\pm$ 10    &     6.7 $\pm$ ~~7.0   \\
to3  &  9000 $\pm$ 300  &  4.1 $\pm$ 0.2    &   0.00$^{a}$~~~~  &   A2\,V$^{b}$   &  133 $\pm$ 10    &     2.8 $\pm$ ~~7.7   \\
to4  &  8300 $\pm$ 400  &  3.5 $\pm$ 0.2    &   0.00$^{a}$~~~~  &   A1\,V$^{b}$   &  199 $\pm$ 20    &     4.0 $\pm$ 10.0  \\
to5  &  9000 $\pm$ 400  &  4.0 $\pm$ 0.2    &   0.00$^{a}$~~~~  &   A1\,V$^{b}$   &  108 $\pm$ 10    &     3.0 $\pm$ ~~9.6   \\  
to6  &  9100 $\pm$ 300  &  4.3 $\pm$ 0.2    &   0.00$^{a}$~~~~  &   A0\,V$^{b}$   &  245 $\pm$ 25    &     9.3 $\pm$ 10.6  \\ % para las TO
to7  &  8800 $\pm$ 400  &  4.5 $\pm$ 0.2    &   0.00$^{a}$~~~~  &   A1\,IV$^{b}$  &  165 $\pm$ 15    &    10.3 $\pm$ ~~1.7   \\
to8  &  8800 $\pm$ 300  &  4.5 $\pm$ 0.2    &   0.00$^{a}$~~~~  &   A1\,V$^{b}$   &   94 $\pm$ 10    &     5.6 $\pm$ ~~5.9   \\
to9  &  8000 $\pm$ 400  &  3.5 $\pm$ 0.2    &   0.00$^{a}$~~~~  &   A3\,V$^{b}$   &  211 $\pm$ 20    &     0.4 $\pm$ 21.5  \\
to10 &  9100 $\pm$ 300  &  4.4 $\pm$ 0.2    &   0.00$^{a}$~~~~  &   A0\,IV$^{b}$  &   83 $\pm$ 10    &     7.4 $\pm$ ~~4.8   \\
to11 &  8800 $\pm$ 300  &  3.9 $\pm$ 0.2    &   0.00$^{a}$~~~~  &   A0\,IV$^{b}$  &   11 $\pm$ ~~5    &     8.3 $\pm$ ~~0.1   \\
to12 &  8800 $\pm$ 400  &  4.0 $\pm$ 0.2    &   0.00$^{a}$~~~~  &   A0\,V$^{b}$   &  228 $\pm$ 20    &     4.5 $\pm$ ~~8.3   \\
to13 &  8500 $\pm$ 400  &  3.6 $\pm$ 0.2    &   0.00$^{a}$~~~~  &   A0\,V$^{b}$   &  236 $\pm$ 25    &     8.2 $\pm$ ~~8.3   \\
to14 &  8800 $\pm$ 300  &  4.5 $\pm$ 0.2    &   0.00$^{a}$~~~~  &   A0\,V$^{b}$   &  144 $\pm$ 14    &     7.5 $\pm$ ~~6.5   \\
\hline
 
\end{tabular}
\end{center}
{\bf Notes.} $^{a}$ Solar ODF adopted. $^{b}$ Spectral types adopted from SIMBAD. $^{c}$ Possible SB2 system. 
\end{table*}

\subsection{Chemical abundances}\label{sec_abund}

In order to calculate the elemental abundances of our (single) targets we made use of the spectral synthesis technique \citep{Catanzaro11,Catanzaro13}, as we already did 
within the SPA project previously \citep{ASCC123}. 
As a starting point, we took the atmospheric parameters obtained with {\scshape rotfit} to compute 1D Local Thermodynamic Equilibrium (LTE) atmospheric models with the {\scshape atlas9} code \citep{Kurucz93a,Kurucz93b}.
Then, we generated the corresponding synthetic spectra by using the radiative transfer code {\scshape synthe} \citep{Kurucz81}. As an optimization code we exploited ad hoc {\scshape idl} routines
based on the {\it amoeba} minimization algorithm
to find the best solution by minimizing the $\chi^2$ of the differences between the synthetic spectra and the observed ones. At this point, to check the validity of the input parameters we let them vary. 
We always found that the best solution is consistent with the {\scshape rotfit} values reported in 
Table~\ref{tab_params}, so we adopted them for the subsequent analysis. Once we checked the parameters we started to determine the abundances. We focused our analysis on 39 spectral regions of
50\,\AA{} each between 4400 and 6800\,\AA{}. In this way we derived the chemical abundances of 22 elements of atomic number up to 56, namely, C, O, Na, Mg, Al, Si, S, Ca, Sc, Ti, V, Cr, Mn,
Fe, Co, Ni, Cu, Zn, Sr, Y, Zr, and Ba.  For the hottest stars, those around the TO point observed with CAOS, it has been impossible to provide reliable abundances. These A-type stars, with effective temperatures 
above 8000\,K, rotate with moderate/high velocities which prevent the analysis of the few spectral lines observed in their spectra. In fact, the bluest part of the spectra is not sufficiently well exposed even 
for classification purposes so we took the spectral types from the SIMBAD database.

Individual abundances for each star are listed, according to the standard notation  $A($X$)= \log\,$[n(X)/n(H)]\,+\,12, in Tables~\ref{abund_ms} and \ref{abund_gig} for MS stars and giants, respectively. Additionally, 
the cluster mean abundances for each element, in terms of [X/H], are reported in Table~\ref{tab_abund}. They have been calculated by means 
of the weighted average of each star, using the individual errors as weight. The abundances are expressed referring to the solar value 
that we obtained by applying the same procedure to a HARPS-N spectrum of Ganymede \citep[see table 5 in][]{ASCC123}.

For what concerns iron, with the exception of the hottest and fast rotating stars for which we can not measure its abundance, we found an average [Fe/H]=$-$0.13$\pm$0.08. This value is slightly lower than that 
derived by using {\scshape rotfit} but still compatible within the errors. In any case, for clarity's sake,
hereinafter we adopted the weighted mean of both values (obtained from {\scshape rotfit} and {\scshape synthe}, respectively)
as the iron content of the cluster, i.e. [Fe/H]=$-$0.07$\pm$0.06.

We find that abundances derived from giants and dwarfs are compatible within the errors for all the elements except for Ba and Sr, which are clearly overabundant in giants 
(0.48 and 0.38\,dex, respectively), and Co, which is only marginally overabundant. For the remaining elements significant discrepancies are not seen. Only for Na, V, and Cu differences are $\geq$\,0.15\,dex, but still consistent with each other.
Stock\,2 shows solar weighted-mean ratios for $\alpha$-elements ([$\alpha$/Fe]=0.04$\pm$0.05, without including the O) and iron-group elements ([X/Fe]=0.03$\pm$0.03)
while for the heaviest elements, without taking into account Sr and Ba, the cluster exhibits a supersolar ratio ([$s$/Fe]=0.17$\pm$0.04).

\begin{table}[ht]
\caption{Average chemical abundances ([X/H]) for Stock\,2 obtained with {\scshape synthe}.} \label{tab_abund}
\begin{center}
\begin{tabular}{lccc}  
\hline\hline
Element  &  Total  &  MS stars  &  Giants \\
\hline
C    &  $-$0.08 $\pm$ 0.05  &  $-$0.08 $\pm$ 0.05  &  \dots  \\
O    &  $-$0.20 $\pm$ 0.03  &  $-$0.20 $\pm$ 0.03  &  \dots  \\
Na   &  $+$0.14 $\pm$ 0.14  &  $+$0.08 $\pm$ 0.14  &  $+$0.23 $\pm$ 0.14  \\
Mg   &  $-$0.20 $\pm$ 0.10  &  $-$0.25 $\pm$ 0.10  &  $-$0.15 $\pm$ 0.10  \\
Al   &  $-$0.13 $\pm$ 0.15  &  $-$0.18 $\pm$ 0.16  &  $-$0.12 $\pm$ 0.15  \\
Si   &  $+$0.05 $\pm$ 0.08  &  $+$0.03 $\pm$ 0.09  &  $+$0.07 $\pm$ 0.09  \\
S    &  $+$0.05 $\pm$ 0.10  &  $+$0.00 $\pm$ 0.11  &  $+$0.14 $\pm$ 0.11  \\
Ca   &  $-$0.04 $\pm$ 0.09  &  $-$0.02 $\pm$ 0.09  &  $-$0.09 $\pm$ 0.10  \\
Sc   &  $+$0.01 $\pm$ 0.13  &  $+$0.00 $\pm$ 0.13  &  $+$0.03 $\pm$ 0.14  \\
Ti   &  $-$0.06 $\pm$ 0.12  &  $-$0.08 $\pm$ 0.12  &  $-$0.01 $\pm$ 0.13  \\
V    &  $+$0.06 $\pm$ 0.10  &  $+$0.14 $\pm$ 0.11  &  $-$0.03 $\pm$ 0.11  \\
Cr   &  $+$0.02 $\pm$ 0.15  &  $-$0.04 $\pm$ 0.15  &  $+$0.09 $\pm$ 0.15  \\
Mn   &  $-$0.07 $\pm$ 0.15  &  $-$0.09 $\pm$ 0.16  &  $-$0.05 $\pm$ 0.15  \\
Fe   &  $-$0.13 $\pm$ 0.08  &  $-$0.15 $\pm$ 0.09  &  $-$0.10 $\pm$ 0.09  \\
Co   &  $+$0.01 $\pm$ 0.05  &  $+$0.08 $\pm$ 0.06  &  $-$0.09 $\pm$ 0.06  \\
Ni   &  $-$0.04 $\pm$ 0.10  &  $-$0.01 $\pm$ 0.10  &  $-$0.07 $\pm$ 0.11  \\
Cu   &  $-$0.22 $\pm$ 0.10  &  $-$0.16 $\pm$ 0.10  &  $-$0.31 $\pm$ 0.11  \\ 
Zn   &  $-$0.16 $\pm$ 0.09  &  $-$0.20 $\pm$ 0.10  &  $-$0.13 $\pm$ 0.09  \\
Sr   &  $+$0.09 $\pm$ 0.15  &  $-$0.02 $\pm$ 0.15  &  $+$0.46 $\pm$ 0.16  \\
Y    &  $+$0.11 $\pm$ 0.04  &  $+$0.12 $\pm$ 0.04  &  $+$0.06 $\pm$ 0.06  \\
Zr   &  $+$0.00 $\pm$ 0.14  &  $+$0.01 $\pm$ 0.15  &  $-$0.01 $\pm$ 0.14  \\
Ba   &  $-$0.11 $\pm$ 0.09  &  $-$0.20 $\pm$ 0.09  &  $+$0.18 $\pm$ 0.09  \\
\hline

\end{tabular}
\end{center}
\end{table}

%%%%%%%%%%%%%%%%%%%%%%%%%%%%%%%%%%%%%%%%%%%%%%%%%%%%%%%%%%%%%%%%%%%%%%%%%%%%%%%%%%%%%%%%%%%%%%%%%%%%%%%%%%%%%%%%%%%%%%%%%%%%%%%%%%%%%%%%%%%%%%%%%%%%%%%

\section{Reddening and SED fitting}\label{sec_redd}

\begin{figure}
  \centering         
  \includegraphics[width=\columnwidth]{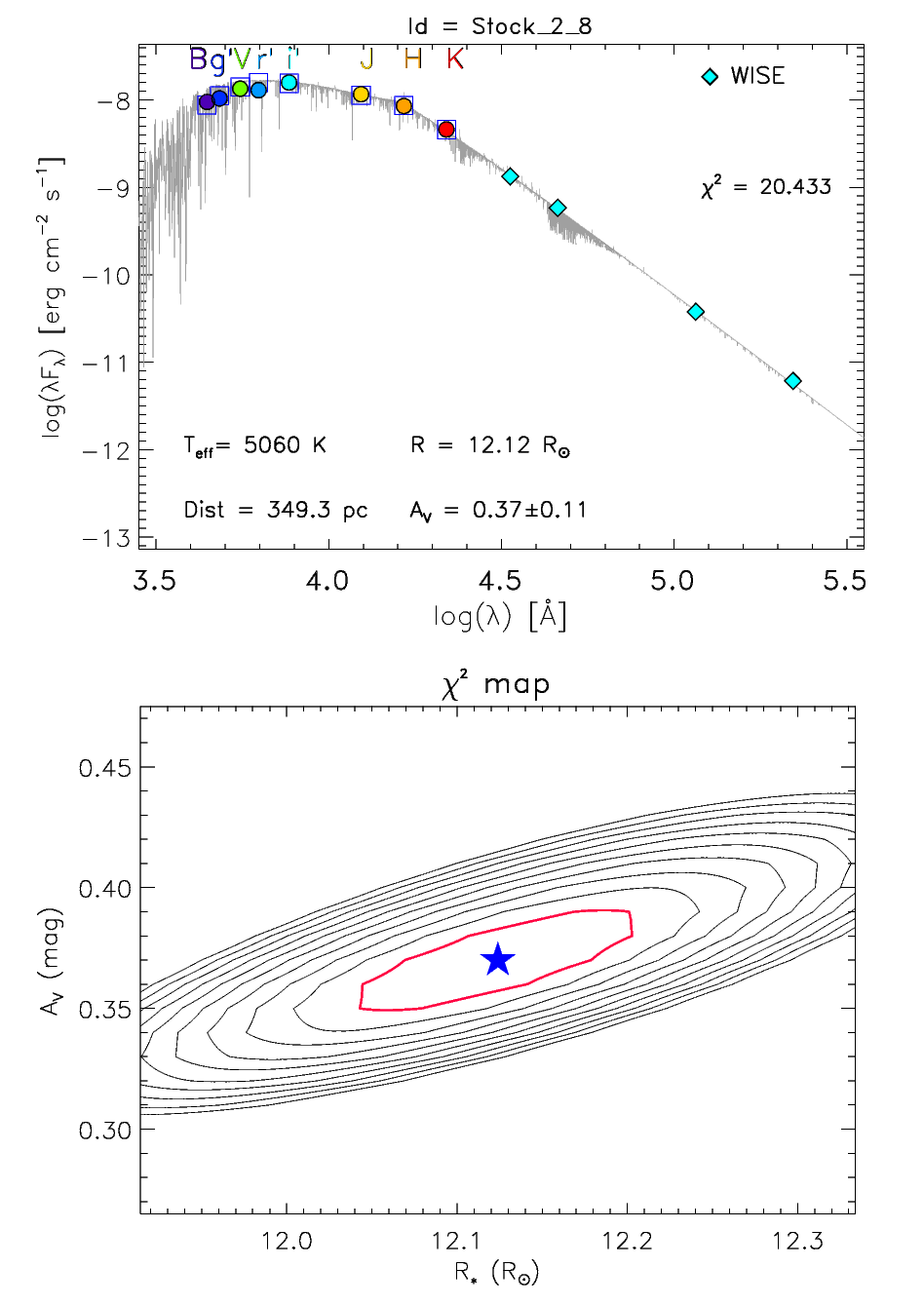}   
  \caption{{\it Top:} Example of a SED fitting (star g8). {\it Bottom:} $\chi^2$-contour map of the fitting. The red contour corresponds to the 1-$\sigma$ 
  confidence level.} 
  \label{fig_sed}
\end{figure}

With the aim of determining the interstellar extinction ($A_V$) of our sources, as well as the luminosity ($L$), we resorted to the spectral energy distribution (SED) 
fitting method. From optical and NIR photometric data publicly available we built the corresponding SED, which was fitted with BT-Settl synthetic spectra \citep{Allard2014}. For each target, we assumed 
its $Gaia$-eDR3 parallax as well as the atmospheric parameters ($T_{\textrm{eff}}$ and $\log\,g$) obtained in 
Sect.~\ref{sec_params}, leaving the stellar radius ($R$) and $A_V$ as free parameters. These parameters were then obtained by $\chi^2$ minimization and the stellar luminosity was calculated as 
$L$=4\,$\pi$\,$R^2$\,$\sigma$\,$T_{\textrm{eff}}^4$. An example of this fitting is shown in Fig.~\ref{fig_sed}.
The errors on $A_V$ and $R$ are found by the minimization procedure considering the 1-$\sigma$ confidence level of the $\chi^2$ map, but we have also taken the error on \teff\ into account .

The $A_V$ values thus obtained are reported in Table~\ref{tab_sed}. In total, we provide results for 42 stars, whose $A_V$ range from 0.37 to 1.93 mag, with an average 
of $A_V$=0.84$\pm$0.34, where the error is the standard deviation. This extinction corresponds to $E(B-V)$=0.27$\pm$0.11 when assuming a standard reddening law with $R_V$=3.1.
The high dispersion confirms the existence of a noticeable differential reddening across the observed field, as described in previous studies \citep{Krzeminski1967,Spagna09}. Indeed, our 
value is compatible within the errors with that mostly accepted for the cluster, $E(B-V)\approx$0.35 \citep{Ye21}.

Alternatively, we evaluated the reddening from the colour excess definition, that is, by comparing observed and intrinsic colours for each star. 
For this purpose we used the 2MASS photometric data shown in Table~\ref{tab_fotom}, since they are more suitable than the optical ones as they are less affected by the extinction.
The intrinsic colours were adopted from the spectral types (Table~\ref{tab_params}) according to the calibrations of \citet{St09}. In this way, from 43 stars, we obtained 
an average cluster reddening of $E(B-V)$=0.26$\pm$0.11, which shows an excellent agreement with the value derived from the SED fitting.
This agreement is especially remarkable considering that photometric calibrations do not take the effect of the rotational velocity on the colour
into account.

\begin{table}
\caption{Results of the SED fitting.\label{tab_sed}}
\begin{center}
\begin{tabular}{lccr} 
\hline\hline
Star  &  $A_V$ (mag) & $R$ (R$_{\sun}$) & \multicolumn{1}{c}{$L$ (L$_{\sun}$)} \\
\hline
b1   &   0.65 $\pm$ 0.17  &  2.23 $\pm$ 0.03  &  23.4 $\pm$ 3.2  \\
b2   &   0.77 $\pm$ 0.08  &  2.12 $\pm$ 0.04  &  23.2 $\pm$ 2.1  \\
b3   &   0.42 $\pm$ 0.22  &  1.85 $\pm$ 0.02  &  10.9 $\pm$ 1.7  \\
b4   &   0.73 $\pm$ 0.24  &  1.80 $\pm$ 0.03  &  10.7 $\pm$ 1.6  \\
b5   &   0.77 $\pm$ 0.25  &  1.64 $\pm$ 0.03  &   6.8 $\pm$ 0.9  \\
b6   &   0.45 $\pm$ 0.10  &  1.15 $\pm$ 0.02  &   1.7 $\pm$ 0.1  \\
b7   &   0.47 $\pm$ 0.09  &  1.08 $\pm$ 0.02  &   1.4 $\pm$ 0.1  \\
b8   &   0.46 $\pm$ 0.11  &  0.98 $\pm$ 0.02  &   1.0 $\pm$ 0.1  \\
     &                    &                   &                  \\
r1   &   0.53 $\pm$ 0.25  &  2.84 $\pm$ 0.05  &  29.7 $\pm$ 4.4  \\
r2   &   0.85 $\pm$ 0.19  &  2.45 $\pm$ 0.03  &  25.7 $\pm$ 3.7  \\
r3   &   1.28 $\pm$ 0.09  &  2.37 $\pm$ 0.03  &  30.2 $\pm$ 4.1  \\
r5   &   1.06 $\pm$ 0.22  &  1.68 $\pm$ 0.03  &   8.5 $\pm$ 1.2  \\
r6   &   0.94 $\pm$ 0.17  &  1.55 $\pm$ 0.02  &   4.8 $\pm$ 0.3  \\
r7   &   0.86 $\pm$ 0.16  &  1.21 $\pm$ 0.02  &   2.1 $\pm$ 0.2  \\
r8   &   0.82 $\pm$ 0.07  &  1.06 $\pm$ 0.01  &   1.4 $\pm$ 0.1  \\ 
     &                    &                   &                  \\
u3   &   1.29 $\pm$ 0.23  &  2.55 $\pm$ 0.04  &  19.6 $\pm$ 3.0  \\
u4   &   1.93 $\pm$ 0.20  &  1.91 $\pm$ 0.03  &  15.5 $\pm$ 2.2  \\
u5   &   1.23 $\pm$ 0.16  &  1.77 $\pm$ 0.02  &   4.9 $\pm$ 0.4  \\
u6   &   1.51 $\pm$ 0.10  &  1.23 $\pm$ 0.02  &   2.5 $\pm$ 0.2  \\   
     &                    &                   &                  \\
g1   &   0.40 $\pm$ 0.30  & 29.84 $\pm$ 1.55  & 337.6 $\pm$ 31.1  \\
g2   &   0.58 $\pm$ 0.31  & 24.85 $\pm$ 0.48  & 285.4 $\pm$ 27.2  \\
g3   &   0.66 $\pm$ 0.27  & 21.27 $\pm$ 0.30  & 242.6 $\pm$ 22.5  \\
g4   &   0.82 $\pm$ 0.26  & 17.36 $\pm$ 0.23  & 166.9 $\pm$ 15.7  \\
g5   &   0.53 $\pm$ 0.13  & 14.99 $\pm$ 0.20  & 132.7 $\pm$  5.9  \\
g6   &   1.15 $\pm$ 0.25  & 18.30 $\pm$ 0.21  & 188.5 $\pm$ 16.6  \\
g7   &   0.85 $\pm$ 0.12  & 15.62 $\pm$ 0.24  & 144.0 $\pm$  6.7  \\ 
g8   &   0.37 $\pm$ 0.11  & 12.12 $\pm$ 0.17  &  86.8 $\pm$  3.9  \\
g9   &   1.43 $\pm$ 0.08  & 15.78 $\pm$ 0.26  & 147.0 $\pm$  6.8  \\
g10  &   0.90 $\pm$ 0.08  & 12.33 $\pm$ 0.13  &  90.5 $\pm$  3.9  \\
     &                    &                   &                    \\
to2  &   0.78 $\pm$ 0.14  &  4.58 $\pm$ 0.05  &  129.3 $\pm$ 16.9  \\
to3  &   0.59 $\pm$ 0.15  &  4.38 $\pm$ 0.08  &  113.3 $\pm$ 15.1  \\
to4  &   0.65 $\pm$ 0.34  &  4.50 $\pm$ 0.09  &   86.4 $\pm$ 16.7  \\
to5  &   1.00 $\pm$ 0.20  &  4.42 $\pm$ 0.09  &  115.2 $\pm$ 20.5  \\
to6  &   0.37 $\pm$ 0.17  &  3.11 $\pm$ 0.07  &   59.8 $\pm$ 7.9   \\
to7  &   1.04 $\pm$ 0.22  &  4.39 $\pm$ 0.06  &  104.0 $\pm$ 18.9  \\
to8  &   0.48 $\pm$ 0.17  &  3.54 $\pm$ 0.05  &   67.6 $\pm$ 9.1   \\
to9  &   0.77 $\pm$ 0.36  &  4.59 $\pm$ 0.08  &   77.8 $\pm$ 15.5  \\
to10 &   0.90 $\pm$ 0.13  &  3.96 $\pm$ 0.07  &   96.9 $\pm$ 12.7  \\
to11 &   1.06 $\pm$ 0.19  &  4.22 $\pm$ 0.12  &   95.9 $\pm$ 13.3  \\
to12 &   0.61 $\pm$ 0.24  &  3.62 $\pm$ 0.10  &   70.7 $\pm$ 12.9  \\
to13 &   0.93 $\pm$ 0.27  &  4.13 $\pm$ 0.11  &   80.2 $\pm$ 15.1  \\
to14 &   1.23 $\pm$ 0.22  &  4.03 $\pm$ 0.15  &   87.8 $\pm$ 12.4  \\
\hline

\end{tabular}
\end{center}
\end{table}

\section{Colour-magnitude diagrams}\label{sec_CMD}
\begin{figure*}
  \centering         
  \includegraphics[width=18cm]{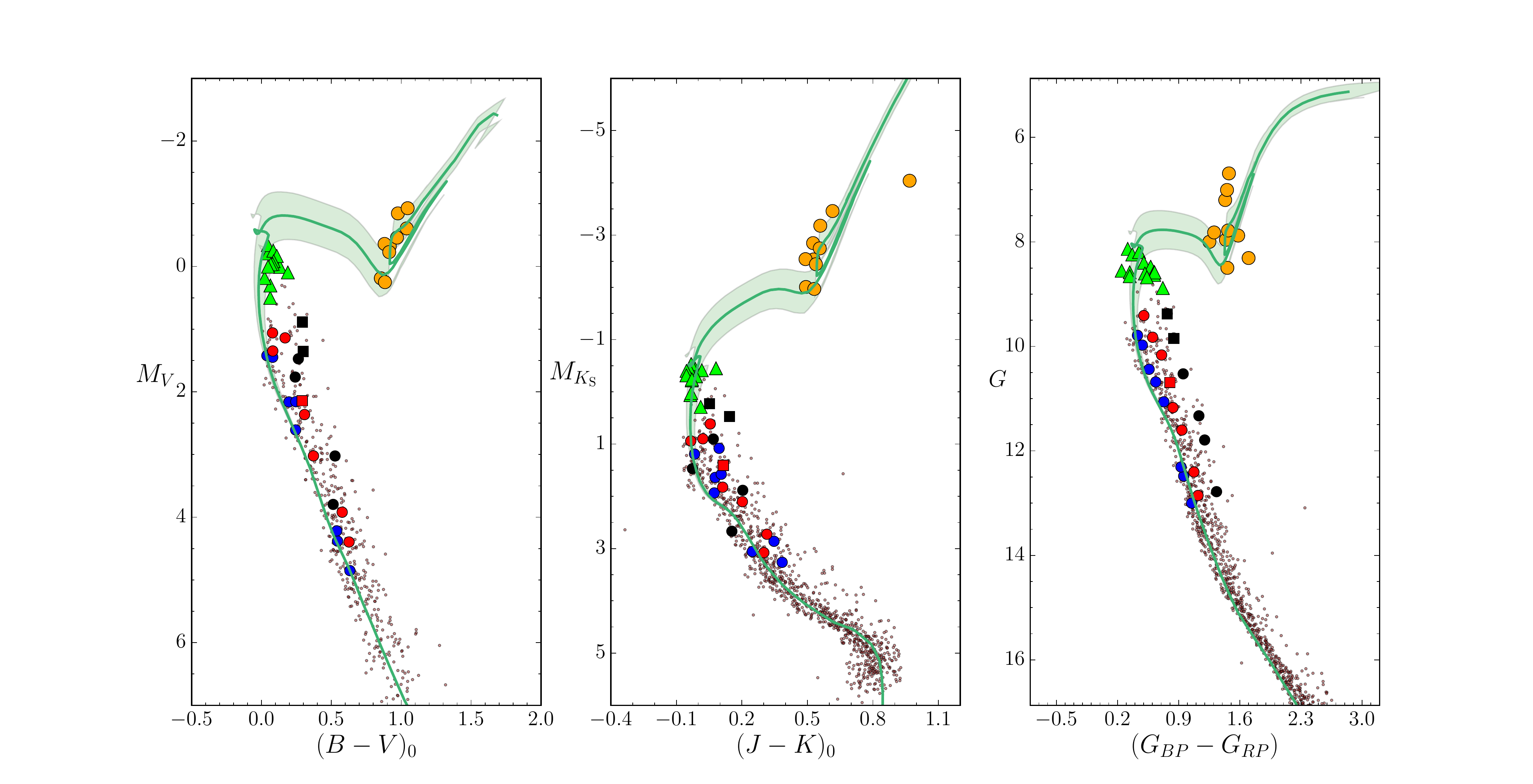}  
  \caption{Colour-magnitudes diagrams for Stock\,2 in three different photometric systems: \textbf{Left:} $M_V$/$(B-V)_0$, photometric data from the APASS 
  catalogue; \textbf{Centre:} $M_{K_{\textrm{S}}}$/$(J-K_{\textrm{S}})_0$ (2MASS) and \textbf{Right:} $G$/$(G_{BP}-G_{RP})$ ($Gaia$-eDR3). Colours and symbols are the 
  same as those in Fig.~\ref{fig_targets}. The green line and the shaded area are the best-fitting isochrone within the uncertainties (log\,$\tau$=8.65$\pm$0.15).} 
  \label{CMD}
\end{figure*}

With the aim of investigating the age of the cluster we combined archival photometry with the spectroscopy obtained in this work.
We made use of the most widespread procedure, the so-called isochrone-fitting method. It consists of finding the age-dependent model, isochrone, that best
reproduces the cluster evolutionary snapshot reflected in its CMD. In a first step, it was necessary to construct the CMD. We did it in three different 
photometric systems (optical, 2MASS and $Gaia$-eDR3), highlighting our targets in Fig.~\ref{CMD}, according to the criterium described in Sect.~\ref{sec_targets}. 
We took advantage of the reddening previously obtained 
($E(B-V)$=0.27, Sect.~\ref{sec_redd}) to draw the following diagrams: $M_V$/$(B-V)_0$, $M_{K_{\textrm{S}}}$/$(J-K_{\textrm{S}})_0$ and $G$/$(G_{BP}-G_{RP})$.
Individual distances, 
derived from the inversion of their parallaxes, were also taken into account. Individual zero-point offset corrections, with an average value around $-$33\,$\mu$as,
were applied to the published $Gaia$-eDR3 parallaxes following the recommendations outlined by \citet{Lindegren21}.
Then, in a second step, we drew PARSEC isochrones \citep{PARSEC} for different ages computed at the metallicity found in this work ([Fe/H]=$-$0.07, see Sect.~\ref{sec_abund}). 
With the intention of ensuring the reliability of the fit, we selected, among the list of members identified by \citet{Cantat2018}, only those with a membership probability 
sufficiently high (i.e. $P\geq$0.7). Additionally, on this sample we imposed a quality cutoff, taking only the objects whose error on parallax is below 0.1 mas, i. e. with an uncertainty 
less than 5$\%$. In total we considered 1016 cluster 
members with $Gaia$-eDR3 photometry. We did a cross-match of our member list with the APASS \citep{apass} and 2MASS \citep{2MASS} catalogues and then we selected 
only the stars with good-quality photometry. In the first case this meant stars with errors on both $V$ and $(B-V)$ <\,0.1 mag, while in the second case, 
just the stars without any `$U$' photometric flag assigned. In total 409 and 955 objects were retrieved, respectively. The resulting diagrams are displayed in 
Fig.~\ref{CMD}.

When building the first CMD ($M_V$/$(B-V)_0$) we immediately realised the wrong position of the brightest stars, among which were many of our targets. 
For these stars the APASS photometry provide errors above one magnitude or even not quantified.
With this purpose we resorted to the ASCC2.5 catalogue \citep{ASCC25} from which we took $V$ and $(B-V)$ for stars brighter than $V$=10, 
after scaling both photometric datasets.\footnote{By employing almost a hundred stars with good-quality photometry in both catalogues, we found average
differences (ASCC2.5 minus APASS) of $\Delta\,V$=$-$0.040 and $\Delta(B-V)$=$-$0.005 mag.} 
Then, we dereddened the CMD (left panel of Fig.~\ref{CMD}) by applying individual corrections to the stars for which we have spectra and the average
value to the rest of stars.
Finally, we plotted the isochrone that best reproduces the CMD based on a visual inspection, from which we obtained for the cluster a log\,$\tau$=8.65$\pm$0.15 (equivalent to an age of 450$\pm$150\,Ma). In this case, the error reflects the interval of 
isochrones that gives a good fit. With this age the MSTO stellar mass is $\approx$2.8\,M$_{\sun}$. In general, stars occupy positions close to the isochrone and only the TO
stars seem to be slightly away from it.

Regarding the 2MASS CMD, the fit is quite good and all stars match the isochrone rather well, with the exception of the star g1. It is the brightest in the cluster and
shows a position away from the the rest of the giants. As it is so bright, it is close to saturate and its photometry, flagged in the catalogue as `$EDD$', has errors in each
band of around 0.2 mag. Therefore, its anomalous $(J-K)$ colour could simply be an instrumental effect. Some residual dispersion is still observed
for the MS stars, although the correction for reddening has been applied; moreover in the NIR the reddening is lower than at optical wavelengths and should play a minor role on the CMD.
After the reddening correction, no clear eMSTO/split MS is apparent in the CMD. Giants show a dispersion in magnitude greater than it would be expected from their
atmospheric parameters, which are very similar to each other. 
 
In the last diagram, the $Gaia$-eDR3 CMD, since the dereddening of the $Gaia$ photometry is not a trivial task, the isochrone (and not the stars as in the previous CMDs)
was reddened using the average extinction obtained in Sect.~\ref{sec_redd}.
A distance modulus of 7.87, which corresponds to the distance derived by \citet{Cantat2018}, was applied.
The fit is also good and stars lie along the isochrone.

%%%%%%%%%%%%%%%%%%%%%%%%%%%%%%%%%%%%%%%%%%%%%%%%%%%%%%%%%%%%%%%%%%%%%%%%%%%%%%%%%%%%%%%%%%%%%%%%%%%%%%%%%%%%%%%%%%%%%%%%%%%%%%%%%%%%%%%%%%%%%%%%%%%
       
\section{Discussion}\label{sec_discuss}

One of the objectives of this research was to determine the age of the cluster.
Now, based on $Gaia$-eDR3 individual parallaxes for the cluster members and the extinction derived from the SED fitting we were able to build 
suitable CMDs, in which the cluster age was obtained via the isochrone-fitting method. 
By analysing the dereddened 2MASS CMD, which is less affected by the interstellar dust than the ones at optical wavelengths used in past works, we can asses that Stock\,2 is a 
moderately young open cluster of 450$\pm$150\,Ma. Therefore, it is somewhat younger than the Hyades and clearly older than the Pleiades. This confirms the results of 
\citet{Spagna09} and \citet{Sciortino2000} over older studies \citep[e.g.][]{Krzeminski1967}.

\begin{figure} 
  \centering         
  \includegraphics[width=\columnwidth]{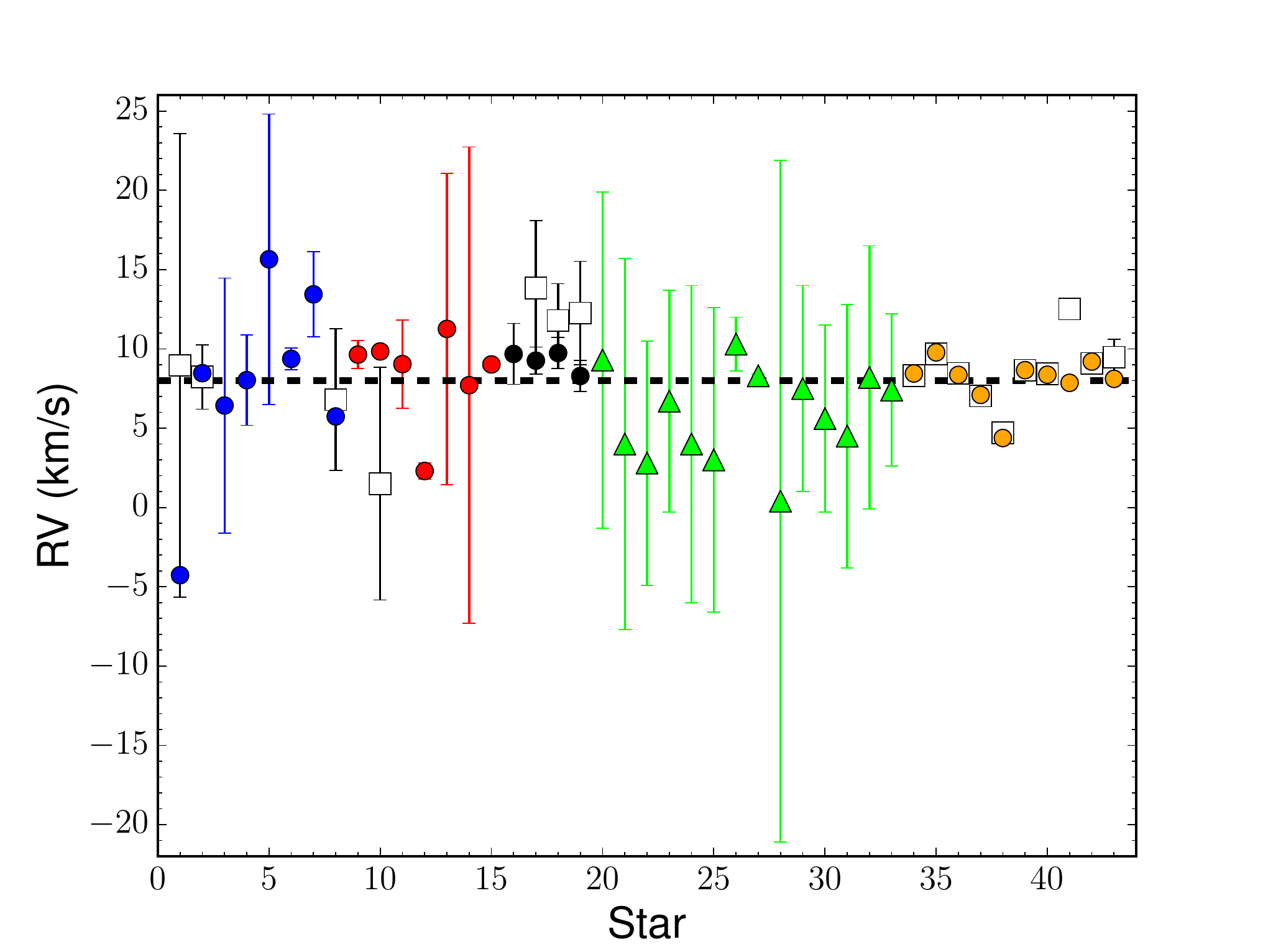} 
  \caption{Comparison of the RVs obtained in this work (symbols and colours as in previous figures) with those of $Gaia$-DR2 (open squares). The dashed line
  shows the cluster average value, $RV$=8.0 km\,s$^{-1}$.} 
  \label{fig_vrad_gaia}
\end{figure}

The RVs obtained by us are, in general, compatible within the errors with those found in the literature, as displayed in Table~\ref{tab_comp_rv} for stars 
in common with \citet{Me08}, who measured RVs for red giants in open clusters, and \citet{Reddy2019}. Although \citet{Me08} claimed binarity for g3 and 
g9, we have not seen any feature in their spectra that might confirm it, as also \citet{Reddy2019} concluded. 
However, given the discrepancies for the latter, perharps it might be a long-period variable.
Figure~\ref{fig_vrad_gaia} shows the 
stars for which we have derived their RV compared, when possible, to the values obtained by $Gaia$-DR2. We remark the excellent agreement for the slow 
rotators, especially in the case of giant stars. For fast rotators, instead, as we already noted, our errors are very large and results are not very reliable; for most 
of them $Gaia$-DR2 does not provide any RV.

\begin{table}
\caption{Comparison of the RV (km\,s$^{-1}$) derived in this work and in the literature.}
\label{tab_comp_rv}
\begin{center}
\begin{tabular}{lcccc}  
\hline\hline
Star  &  Me08 & $Gaia$-DR2  &  Reddy19  &  This work  \\  
\hline
g3  &  9.6 $\pm$ 1.3  &  8.5 $\pm$ 0.2  &  8.8 $\pm$ 0.1  &  8.4 $\pm$ 0.1  \\
g4  &  8.1 $\pm$ 0.4  &  8.4 $\pm$ 0.1  &  8.6 $\pm$ 0.1  &  8.4 $\pm$ 0.1  \\
g9  &  7.8 $\pm$ 0.8  &  4.7 $\pm$ 0.2  &  4.4 $\pm$ 0.1  &  4.4 $\pm$ 0.1  \\
\hline 
 
\end{tabular}
\end{center}
\end{table}

Regarding the atmospheric parameters, as already mentioned, \citet{Reddy2019} conducted the only spectroscopy-based paper devoted to Stock\,2. Their study is 
based on high-resolution spectra ($R$=60\,000) of three of the cluster giants. These stars, which have also been observed by us, are g3 (numbered as 43 in their work), 
g4 (1011) and g9 (1082). Our temperatures and metallicities are slightly larger but still in agreement with their values, within the errors.
Instead, gravities are only marginally compatible. Both datasets are compared in Table~\ref{tab_comp_par}. These discrepancies probably can be explained because of the 
different methodology followed. In this work we employed spectral fitting while their approach was based on the equivalent width (EW) analysis.
       
\begin{table*}[ht]
\caption{Comparison of the atmospheric parameters derived in this work with those of the literature. }
\label{tab_comp_par}
\begin{center}
\begin{tabular}{l|ccc|ccc} 
\hline\hline
\multirow{2}{*}{Star}      &     \multicolumn{3}{c}{This work}            &       \multicolumn{3}{c}{\citet{Reddy2019}}     \\ 
                           & $T_{\textrm{eff}}$ (K) & $\log\,g$ & [Fe/H]     &     $T_{\textrm{eff}}$ (K) & $\log\,g$ & [Fe/H]  \\  
\hline
g3  &  4937 $\pm$ 114  &  2.51 $\pm$ 0.35  &  0.04 $\pm$ 0.08  &  4925 $\pm$ 50  &  2.0 $\pm$ 0.1  &  $-$0.07 $\pm$ 0.03  \\
g4  &  4977 $\pm$ 117  &  2.82 $\pm$ 0.35  &  0.04 $\pm$ 0.08  &  4900 $\pm$ 50  &  2.3 $\pm$ 0.1  &  $-$0.05 $\pm$ 0.04  \\
g9  &  5062 $\pm$ 56   &  2.98 $\pm$ 0.18  &  0.00 $\pm$ 0.09  &  5050 $\pm$ 50  &  2.6 $\pm$ 0.1  &  $-$0.06 $\pm$ 0.03  \\
\hline 
 
\end{tabular}
\end{center}
\end{table*}
     
\begin{figure}
  \centering         

  \includegraphics[width=7cm]{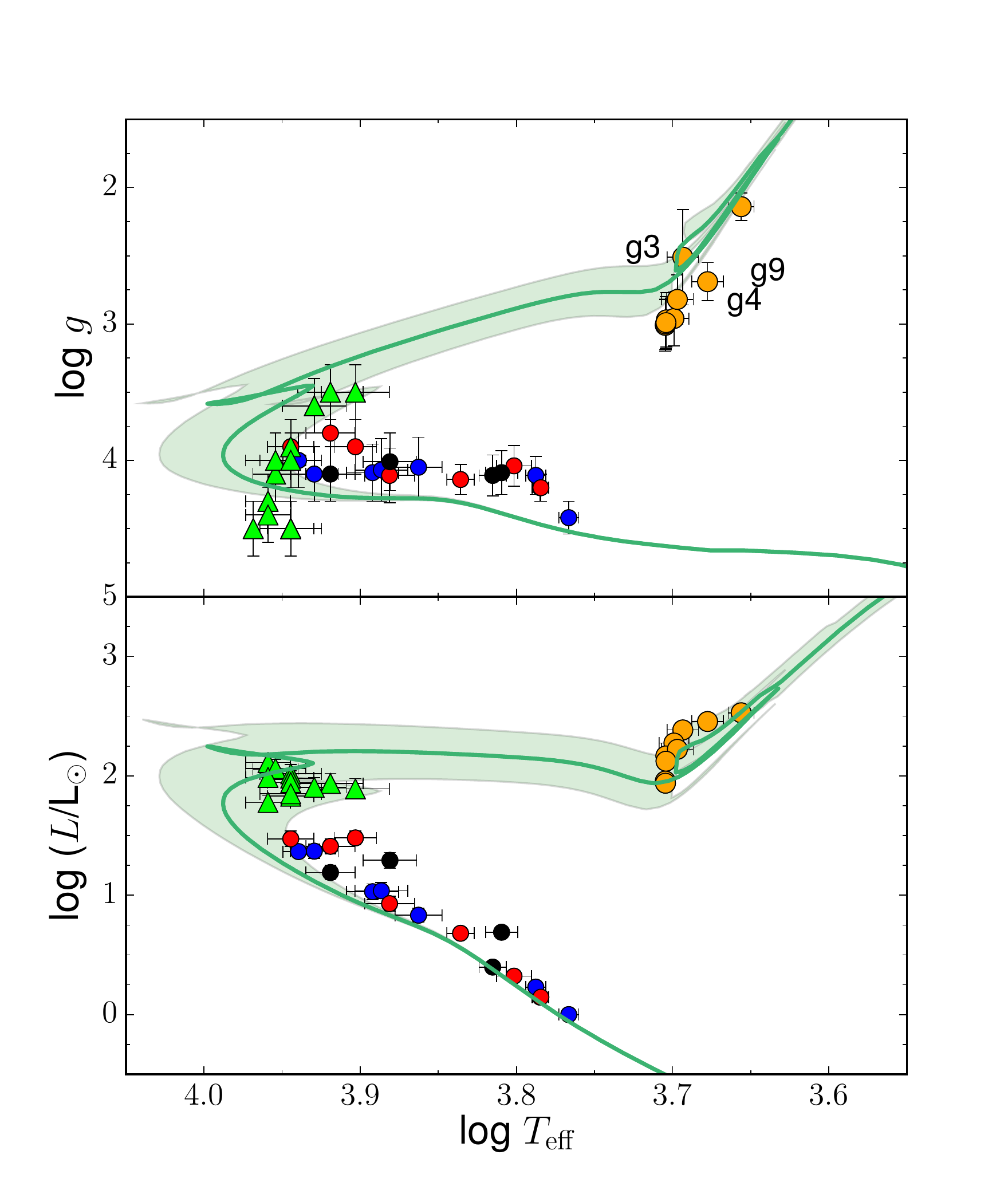}  
  \caption{Kiel and HR diagrams for Stock\,2. Symbols and colours are the same as those in Fig.~\ref{CMD}.} 
  \label{fig_phr} 
\end{figure}

With the aim of checking the consistency of our results, we plot the Kiel and HR diagrams in Fig.~\ref{fig_phr}. The former is a reddening-free diagnostic 
whereas in the latter, extinction has been taken into account when calculating the luminosity. The location of the stars in the HR diagram is better than in the
Kiel diagram, where gravities lie away with respect to those of the isochrone around 0.2 dex, as already came out in the comparison with results from \citet{Reddy2019}. Additionally, 
TO stars show a large dispersion in this diagram. This is very likely a consequence of the poor accuracy of the gravity determinations for these A-type stars, 
which have a moderate or fast rotation.
On the contrary, in the HR diagram these stars are placed more closely clustered around the TO point, as it is expected. 
The fit is also better for MS stars and especially good for giants, which fall on the isochrone. 
       
\subsection{Chromospheric emission and lithium abundance}
\label{Sec:chrom_lithium}

\begin{figure}
\begin{center}
\hspace{-.5cm}
\includegraphics[width=9.0cm]{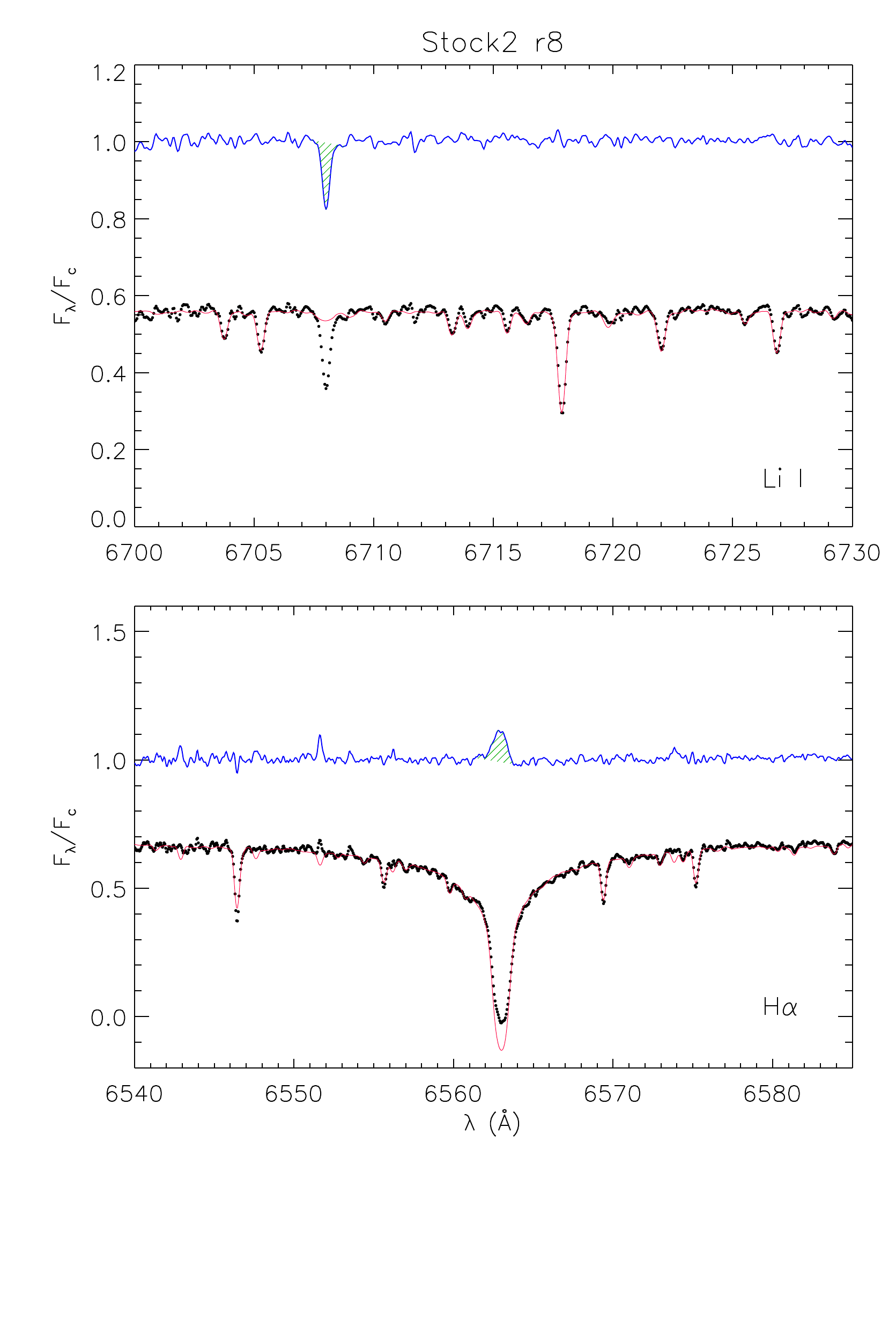}
\vspace{-1.5cm}
\caption{Subtraction of the non-active, lithium-poor template (red line) from the spectrum of Stock2~r8 (black dots), which reveals the chromospheric emission in the H$\alpha$ core (blue
line in the {\it bottom panel}) and emphasizes the \ion{Li}{i} $\lambda$6708\,\AA\ absorption line,
removing the nearby blended lines ({\it top panel}). The green hatched areas represent the excess H$\alpha$ 
emission ({\it bottom panel}) and \ion{Li}{i} absorption ({\it top panel}) that were  integrated to
obtain $EW_{\rm H\alpha}^{em}$ and $EW_{\rm Li}$, respectively.}
\label{fig:subtraction}
\end{center}
\end{figure}

For stars cooler than about 6500\,K and with an age from a few ten to a few hundred Ma, the level of magnetic activity (e.g. the emission in the cores of lines formed in the chromosphere) 
and the atmospheric lithium abundance can be used to estimate the age \citep[see, e.g.,][and references therein]{Jeffries2014, Frasca2018}.
The best diagnostics of chromospheric emission in the wavelength range covered by HARPS-N are \ion{Ca}{ii} H\&K and Balmer H$\alpha$ lines.
However, the S/N ratio at 3900\,\AA\ is very low, so that we can only use the H$\alpha$ for this purpose.
The templates produced by {\scshape rotfit} with rotationally broadened spectra of non-active, lithium-poor stars were 
subtracted from the observed spectra of the targets to measure the excess emission in the core of the H$\alpha$ line ($EW_{\rm H\alpha}^{em}$) and the equivalent 
width of the \ion{Li}{i} $\lambda$6708\,\AA\ absorption line ($EW_{\rm Li}$), removing the blends with nearby lines.

\setlength{\tabcolsep}{5pt}

\begin{table}[htb]
\caption{H$\alpha$, \ion{Li}{i}$\lambda$6708\,\AA\ equivalent widths and lithium abundance for the targets cooler than 7000\,K.}
\begin{center}
\begin{tabular}{lcrrrrl}
\hline
\hline
\noalign{\smallskip}
Star               & \teff &  $EW_{\rm H\alpha}^{em}$    & err   & $EW_{\rm Li}$ & err & $A$(Li)    \\  
                   & (K)   &  \multicolumn{2}{c}{(m\AA)}    & \multicolumn{2}{c}{(m\AA)}  & {(dex)}   \\  
\hline
\noalign{\smallskip}
 b6  &  6132 & 143   &   24  &   63 &   6   &  $2.68^{+0.10}_{-0.10}$ \\
 b7  &  6092 & $\dots$ & $\dots$ &  <3  & $\dots$ &  $<$1.27              \\ 
 b8  &  5841 &  72   &   31  &  145 &  12   &  $2.93^{+0.11}_{-0.10}$ \\ 
 r6  &  6851 & $\dots$ & $\dots$ &   54 &   5   &  $3.03^{+0.12}_{-0.13}$ \\
 r7  &  6332 &  50  &    15  &    9 &   6   &  $1.91^{+0.31}_{-0.56}$ \\
 r8  &  6086 & 110  &    17  &   89 &  10   &  $2.88^{+0.10}_{-0.11}$ \\
 u5  &  6449 &  38  &    13  &   32 &   6   &  $2.55^{+0.18}_{-0.20}$ \\ 
 u6  &  6534 &  93  &    37  &   15 &  10   &  $2.23^{+0.32}_{-0.55}$ \\
\hline
\end{tabular}
\end{center}
\label{Tab:Halpha_Lithium}
\end{table}

Figure~\ref{fig:subtraction} shows an example of the subtraction procedure used to measure the equivalent width of H$\alpha$ and lithium lines, 
$EW_{\rm H\alpha}^{em}$ and $EW_{\rm Li}$. These quantities were measured on the subtracted spectra  by integrating the residual emission and absorption
profiles, as shown by the green dashed areas in Fig.~\ref{fig:subtraction}, and are reported in Table~\ref{Tab:Halpha_Lithium}. 

\begin{figure}[]
\includegraphics[width=9.0cm]{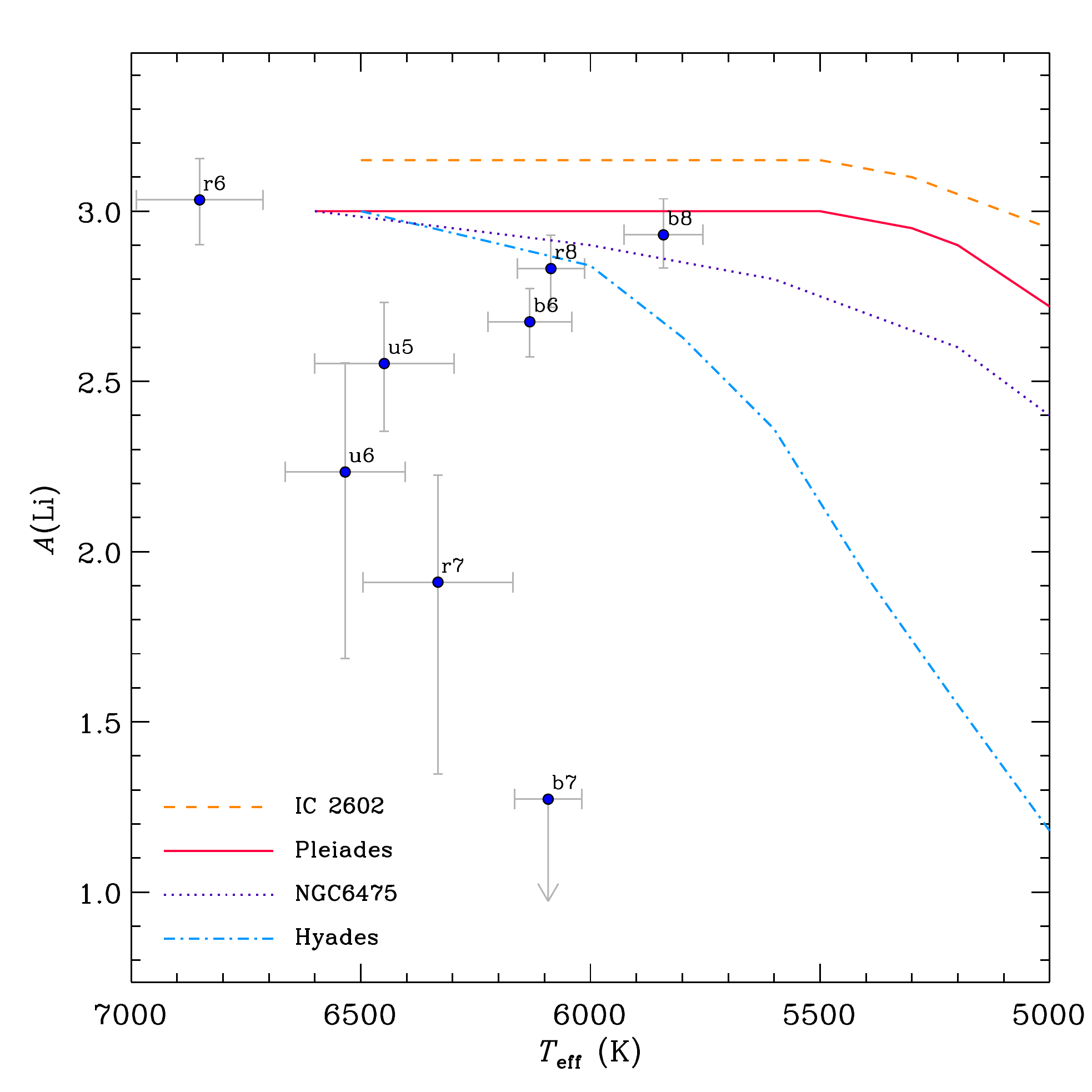}           
\caption{Lithium abundance as a function of $T_{\rm eff}$. 
The upper envelopes of $A$(Li) for IC~2602 ($age\approx$\,30 Ma), Pleiades ($\approx$\,100 Ma),  NGC\,6475 ($\approx$\,300 Ma), and Hyades ($\approx$\,600 Ma) clusters adapted from \citet{Sestito2005} are overplotted.
}
\label{Fig:NLi}
\end{figure}

A simple method to get an estimate of a star's age independent of that derived from isochrones is to compare its position in a diagram that plots lithium abundance, 
$A$(Li), versus \teff\ with the upper envelopes of clusters with a known age. 
We calculated the lithium abundance, $A$(Li), from our values of \teff, \logg, and $EW_{\rm Li}$ by interpolating the curves of growth of \citet{Lind2009}, which span 
the \teff\ range 4000--8000\,K and \logg\ from 1.0 to 5.0 and include non-LTE corrections. In Fig.\,\ref{Fig:NLi} we show the lithium abundance as a function of $T_{\rm eff}$ 
along with the upper envelopes of the distributions of some young open clusters shown by \citet{Sestito2005}. Apart from the large errors of $A$(Li), which take into account both 
the $T_{\rm eff}$ and $EW_{\rm Li}$ errors,
Fig.\,\ref{Fig:NLi} shows that all the targets are located close or below the Hyades upper envelope, compatible with an age $\approx$\,600\,Ma. The only exception is the coldest 
target, b8, which lies between the upper envelopes of the Pleiades ($\approx$\,100\,Ma) and NGC\,6475 ($\approx$\,300\,Ma), which suggests an age $\lesssim 300$\,Ma for this star. 
However, for stars with $T_{\rm eff}$>6000\,K the upper envelopes are very close to each other, which hampers the estimation of the cluster's age with this method.
Lithium abundances for colder stars, where the envelopes separate more, would be extremely useful in clarifying this point. Unfortunately, the combination of very high resolution and telescope size 
did not permit to reach the low main sequence. Hopefully, large samples of fainter stars will be acquired, e.g. by the survey WEAVE \citep{Dalton20} due to start soon 
at the 4.2-m William Herschel Telescope.

\subsection{Galactic metallicity gradient}      

\begin{figure}[h!]
  \centering         
  \includegraphics[width=\columnwidth]{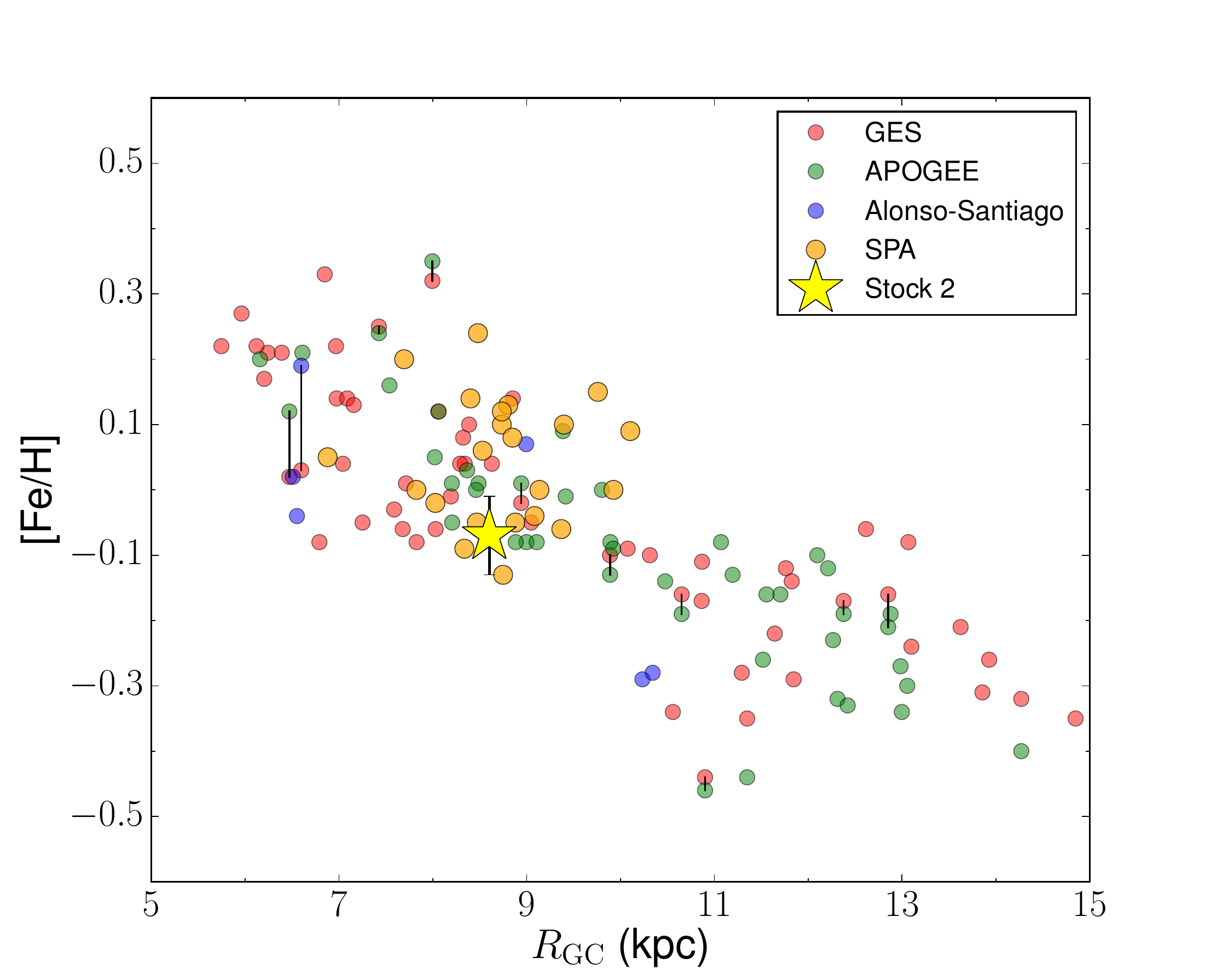} 
  \caption{Radial metallicity gradient from open clusters studied in the framework of the $Gaia$-ESO \citep[][red circles]{Baratella20, Magrini21} and 
  APOGEE \citep[][green circles]{Donor20} surveys. Other similar clusters analysed by \citet[][blue circles]{6067, 3105, 2345, 3OC} are also added along to those
  previously investigated in the SPA project \citep[][orange circles]{ASCC123, D'Orazi20, Casali20, Zhang21}. Black lines link 
  results for the same cluster provided by different authors. The star-symbol represents Stock\,2.} 
  \label{grad}
\end{figure}

Open clusters are good tracers of the radial metallicity distribution of the Galaxy (i.e. the so-called Galactic gradient).
To see how the metallicity derived for Stock\,2 in this work compares with the general gradient, we collected a sample of homogeneously
analysed clusters from the $Gaia$-ESO iDR5 and iDR6 \citep{Baratella20,Magrini21} and the APOGEE DR16 surveys \citep{Donor20}. From 
the latter we only took clusters with data derived from two or more stars and closer than 15 kpc. In addition, open clusters from 
\citet{6067,3105,2345,3OC} are also added to the sample along with those previously investigated within the SPA project \citep{ASCC123, D'Orazi20, Casali20, Zhang21}.
In total, for this comparison we gathered more than a hundred clusters, ten of which are in common among 
different datasets. Figure~\ref{grad} shows the location of Stock\,2 in the Galactic gradient. Galactocentric distances have been taken from \citet{Cantat2018}, 
which obtained their distances from the $Gaia$-DR2 parallaxes, taking as a reference for the solar value $R_{\sun}$=8.34\,kpc. The metallicity, in terms of iron 
abundance, was referenced to $A$(Fe)=7.45\,dex \citep{Grevesse07}. The metallicity found in this work is compatible with that expected for its 
position.

\subsection{Chemical composition and Galactic trends}       
       
\begin{figure} 
  \centering         

  \includegraphics[width=\columnwidth]{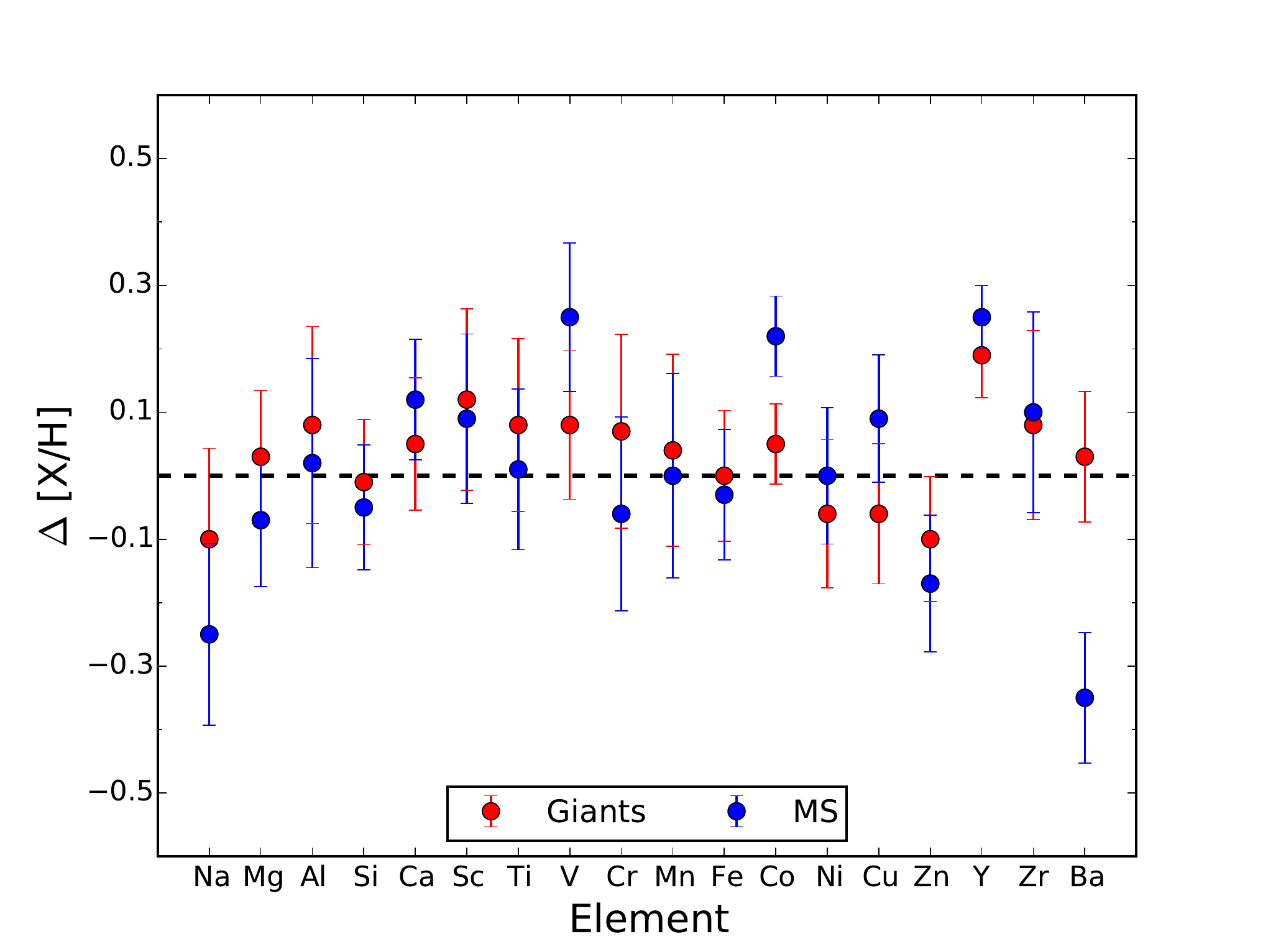} 
  \caption{Differences between our mean abundances, for giants and MS stars, and those by \citet{Reddy2019}. The error bars are the quadratic sum of the
  uncertainties reported in both studies for each element.} 
  \label{comp_reddy} 
\end{figure}

Regarding the abundances, we compared our results (separately for MS stars and giants) to those of \citet{Reddy2019}, with which we have 17 chemical elements in common. 
For the comparison, the values from \citet{Reddy2019} have been scaled to our solar references. In Fig.~\ref{comp_reddy} the differences of the abundance ratios ([X/H]),
this work minus literature, are displayed. As expected, differences are smaller for giants (on average, $\Delta$[X/H]=0.07\,dex) than for MS stars (0.12 dex).
With the only exception of Y, the chemical composition of all the giants is fully compatible with that obtained by \citet{Reddy2019}. On the other hand, 
for MS stars, abundances for Na, V, Co, Zn, Y and Ba are somewhat different.

\begin{figure*}[ht]
  \centering         
  \includegraphics[width=16cm]{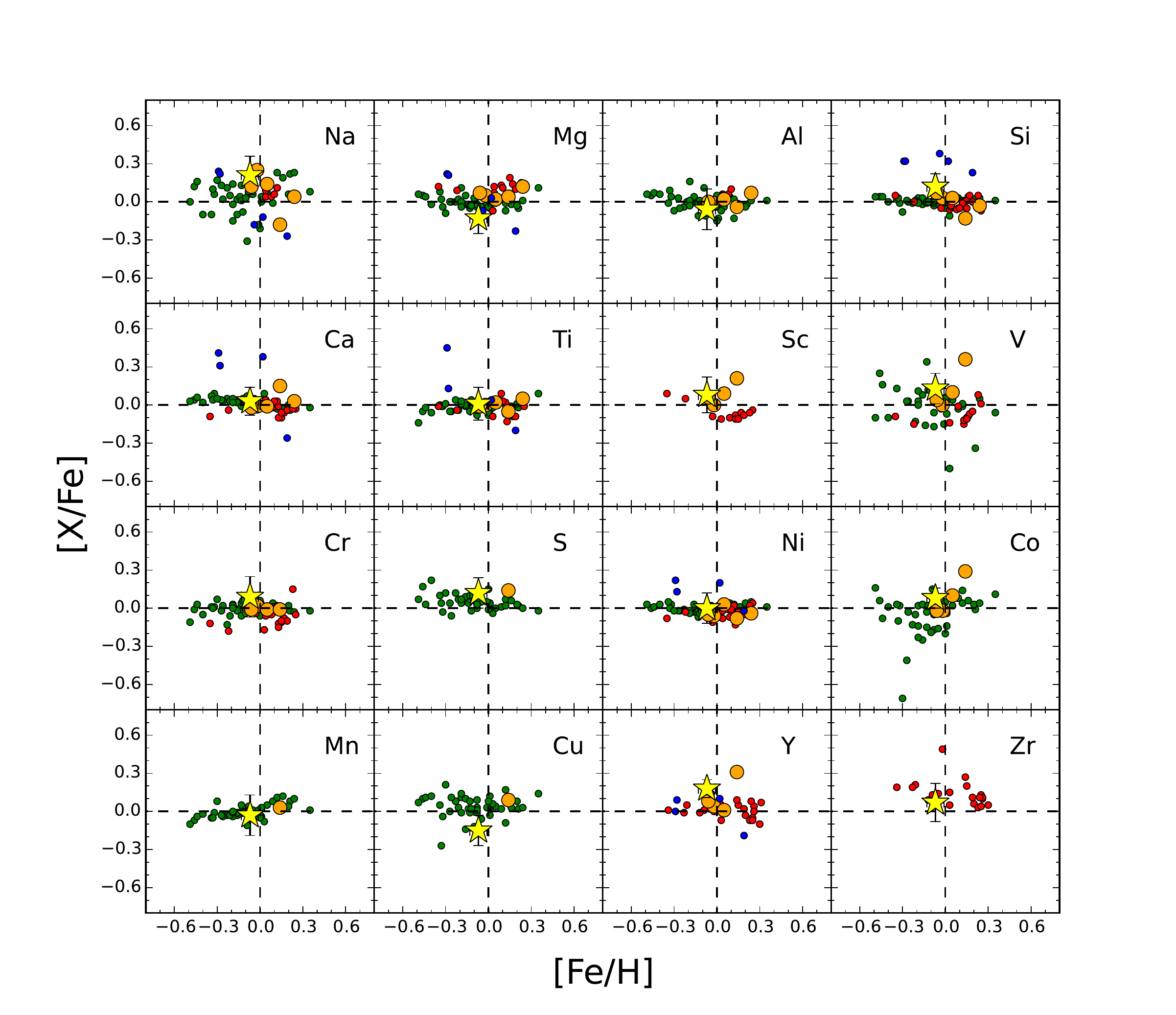} 
  \caption{Abundance ratios [X/Fe] vs. [Fe/H]. Symbols and colours are the same as in Fig.~\ref{grad}. The dashed lines indicate the solar value.} 
  \label{trends}
\end{figure*}

Finally, as we have done above in relation to the metallicity gradient, we contrast the abundances obtained in this work with those of the 
comparison clusters selected before. We completed the sample by adding the $Gaia$-ESO DR4 abundances \citep{Magrini17, Magrini18} for the clusters in 
common with \citet{Magrini21}.
In total, we have in common with them up to 18 chemical elements, out of which the ratios [X/Fe] versus [Fe/H] are displayed in Fig.~\ref{trends} for 16 chemical elements.
The remaining two are O and Ba but since for these elements, the measure of the abundances is conditioned by the evolutionary state of the stars (see Sect.~\ref{sec_abund}), we discarded them from the comparison.
In general, Stock\,2 shows a chemical composition compatible with that of the Galactic thin disc, as supported by the agreement with the observed chemical trends
traced by more than a hundred open clusters. Only the abundance of Cu is sligthtly below these trends, but it is still compatible with them.

\subsection{Rotational velocity, reddening and eMSTO}

We investigated the relationship between $v$\,sin\,$i$ and the eMSTO phenomenon. As mentioned in Sect.~\ref{sec_targets}, we selected our targets following 
in the CMD of Fig.~\ref{fig_targets} three different sequences along the MS: blue, red and the upper envelope. Among all the stars observed
in this work around 40$\%$ rotate rapidly (with $v$\,sin\,$i$\,>\,100\,km\,s$^{-1}$). As can be seen in Table~\ref{tab_params}, in general, the 
fastest rotators are found among the brightest stars in each sequence but also a large scatter of velocities is detected. 
According to the literature \citep{Dupree17,Marino18b,Sun19} the bMS should be populated by stars that rotate slower than those in the rMS.
However, this is not what we observe in this work. Significant differences are not found in 
the mean $v$\,sin\,$i$ of both sequences. In addition, for those single stars in the group in which we expected to find binaries (the upper envelope sequence), 
their $v$\,sin\,$i$ are smaller than in the two other series, despite being redder even than the rMS stars (see Table~\ref{tab_vsini}).

\setlength{\tabcolsep}{7pt}

\begin{table}
\caption{Mean projected rotational velocities (km\,s$^{-1}$) and reddening along MS stars. N is the number of stars in each category.}
\label{tab_vsini}
\begin{center}
\begin{tabular}{lrc}   %longtable es para poner primero landscape
\hline\hline
\multirow{2}{*}{MS sequence (N)}  & \multicolumn{1}{c}{$v$\,sin\,$i$}   & $A_V$     \\
                 &  (km\,s$^{-1}$)   &   (mag)  \\
\hline
bMS (8)   &  103 $\pm$ 106 &  0.59 $\pm$ 0.15  \\ 
rMS (7)   &  100 $\pm$ ~~98  &  0.91 $\pm$ 0.23  \\
uMS (4)   &   57 $\pm$ ~~22  &  1.49 $\pm$ 0.32  \\

\hline 
 
\end{tabular}
\end{center}
\end{table}

\setlength{\tabcolsep}{5pt}

\begin{figure} 
  \centering         
  \includegraphics[width=\columnwidth]{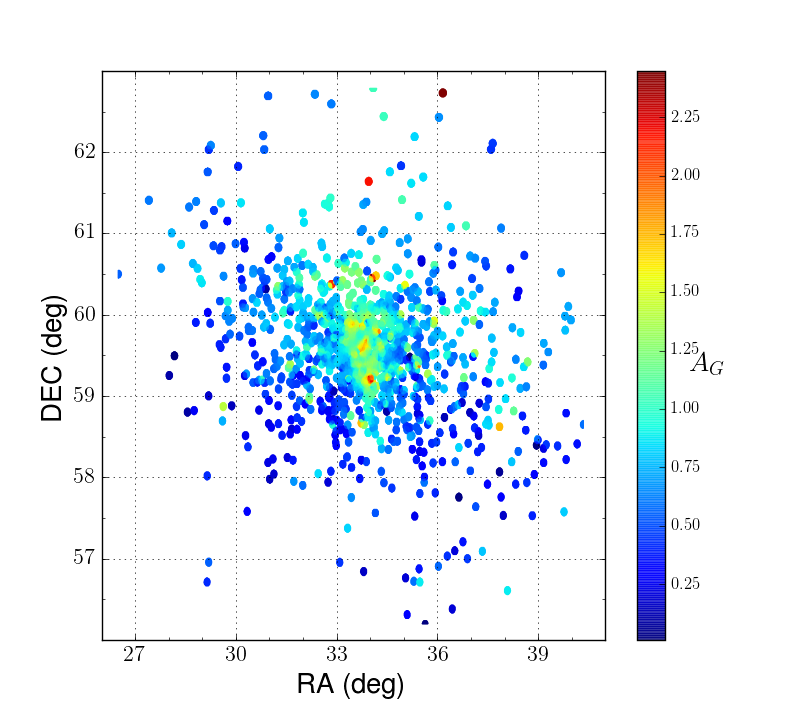}   
  \caption{Interstellar extinction ($A_G$) towards Stock\,2, as traced by the cluster members.} 
  \label{fig_extinct} 
\end{figure}

To interpret this phenonomenon the contribution of the reddening should not be ignored. The cluster average value obtained in this work is compatible within
the errors with that expected for its position according to the extinction maps obtained by \citet{Lallement19}. However, as noted above, its value varies
considerably across the cluster field. For illustrative purposes only, in Fig.~\ref{fig_extinct} we mapped the distribution of $A_G$ in the cluster region 
from its members. 
Since $Gaia$-eDR3 does not provide these values, we took them from $Gaia$-DR2. For slightly more than half of the members identified by \citet{Cantat2018}, 
specifically for 673 stars, their $A_G$ were available. In order to derive individual values for the remaining objects we calculated them as the distance-weighted 
average of the values of the five closest members. Once we estimated the $A_G$ for all the members we started to construct the chart. In a first step, a grid of points
covering the spatial distribution of the cluster members was generated. These points were spaced every 30$\arcsec$ in both RA and DEC. Then, in a second step, the $A_G$ 
of all the members distant up to 3$\arcmin$ from each point was averaged. The resulting spatial distribution of the cluster members, colour-coded according 
to their $A_G$, is shown in Fig.~\ref{fig_extinct}. It displays how variable is the 
reddening across the cluster field, which is likely the result of the low Galactic latitude and the large extension that it occupies in the sky. 

For each of the sequences in which we grouped our MS stars, we calculated the average $v$\,sin\,$i$ and $A_V$. 
These quantities, together with their standard deviations, are quoted in Table~\ref{tab_vsini}. Although our sample is not statistically large, our data 
suggest that rotational velocity cannot explain the observed eMSTO, while the reddening is the most likely responsible for it.

%%%%%%%%%%%%%%%%%%%%%%%%%%%%%%%%%%%%%%%%%%%%%%%%%%%%%%%%%%%%%%%%%%%%%%%%%%%%%%%%%%%%%%%%%%%%%%%%%%%%%%%%%%%%%%%%%%%%%%%%%%%%%%%%%%%%%%%%%%%%%%%%%%%%%%%%%

\section{Conclusions}\label{sec_concl}

We have conducted this research in the framework of the SPA project with the aim of continuing to improve our knowlegde of the solar neighbourhood.
This work is focused on Stock\,2, a nearby and little-studied open cluster. We performed its detailed study from high-resolution spectroscopy complemented 
with archival photometry and $Gaia$-eDR3 data. Our sample, by far the largest to date, is composed of 46 bona-fide members, including both giants
and MS stars. Among the latter, in order to study the eMSTO phenomenon, we selected the brightest stars around the TO point and many others following three
different sequences to cover the spread observed in the CMDs.

We found three double spectrum binaries in our sample. For the rest of the stars we measured their radial and projected rotational velocities and derived the extinction and 
their atmospheric parameters. In addition, we carried out the chemical analysis for 29 stars observed with HARPS-N providing the abundances of 22 elements.

We found that half of the MS stars are fast rotators, with $v$\,sin\,$i$>100\,km\,s$^{-1}$. However, the distribution of slow and fast rotators along the 
bMS, rMS and uMS sequences is random, which discards the rotational velocity as the cause of the observed eMSTO. Additionally, cluster members 
are disseminated over a wide region of the sky (up to $\approx$13$^{\circ}$\,$\times$\,8$^{\circ}$) and differential reddening plays an important role in shaping 
the CMDs. We found an average reddening in the cluster field of $E(B-V)$=0.27$\pm$0.11. Its large dispersion (consistent with the $Gaia$-DR2 value, 
$E(G_{\textrm{BP}}-G_{\textrm{RP}})$=0.40$\pm$0.18) confirms the existence of a variable reddening across the field of Stock\,2.

The reddening also makes it difficult to obtain an accurate age for the cluster. However, from the isochrone-fitting on the dereddened 2MASS CMD, which is the one 
less affected by the extintcion, we derived a value of 450$\pm$150\,Ma. This age implies a mass at the MSTO of $\approx$2.8\,M$_{\sun}$. 
The analysis of the abundance of lithium indicates an age similar to the Hyades ($\sim 600$\,Ma), although the coolest observed member could be as young as 300\,Ma. 
Spectroscopic observations of a larger sample of members with a lower \teff\ is needed to settle this point. We expect very useful data from large spectroscopic surveys 
that will start in the near future, such as WEAVE. The cluster RV derived from the giants is $\approx$8.0\,km\,s$^{-1}$. Stock\,2 shows a solar-like metallicity, [Fe/H]=$-$0.07$\pm$0.06, 
fully compatible within the errors with that expected for its Galactocentric distance.

Finally, we performed a detailed study of the cluster chemical composition by determining the abundances of C, odd-Z elements (Na, Al), $\alpha$-elements (O, 
Mg, Si, S, Ca, Ti), iron-peak elements (Sc, V, Cr, Mn, Co, Ni, Cu, Zn) and $s$-elements (Sr, Y, Zr, Ba). MS stars exhibit a chemical composition 
compatible within the errors with the giants. Only for Co and particularly for Ba and Sr diferences are significant, being the abundances of Ba and Sr clearly higher in giants.
We conclude our research claiming the consistency of its chemical composition with that of the thin disc. This is supported by the values of its ratios [X/Fe] that are on 
the Galactic trends displayed by open clusters in the $Gaia$-ESO and APOGEE surveys. Finally, the cluster shows solar-like mean ratios for the $\alpha$ ([$\alpha$/Fe]=0.04$\pm$0.05) and
the iron-peak [iron-peak/Fe]=0.03$\pm$0.03 elements while for the heaviest elements (without including the Ba and Sr abundances) exhibits a mild overabundance 
with respect to the Sun, [$s$/Fe]=0.17$\pm0.04$.

%%%%%%%%%%%%%%%%%%%%%%%%%%%%%%%%%%%%%%%%%%%%%%%%%%


\bibliographystyle{aa}
\bibliography{Stock2} % if your bibtex file is called example.bib

%%%%%%%%%%%%%%%%% APPENDICES %%%%%%%%%%%%%%%%%%%%%

\appendix

\section{Additional material}\label{appendix}

\begin{table*}
\caption{$Gaia$-eDR3 astrometric data and distance from the nominal cluster centre for the stars observed spectroscopically in this work.}
\label{tab_mp}
\begin{center}
\begin{tabular}{lccccccc}  
\hline\hline
\multirow{2}{*}{Star}  & \multirow{2}{*}{$Gaia$ ID} &  RA     &   DEC   &     $r$     &  $\mu_{\alpha*}$  & $\mu_{\delta}$    & $\varpi$    \\
                       &                                & (J2000) & (J2000) & ($\arcmin$) & (mas\,yr$^{-1}$)  &  (mas\,yr$^{-1}$) &  (mas) \\
\hline
b1    &  459178703132426240  &  34.48311854134  &  58.90486243904  &  31.3  &  14.999  &  $-$14.611  &  2.6447  \\ 
b2    &  507116585468397056  &  32.54413370951  &  59.54252695750  &  40.4  &  16.301  &  $-$13.214  &  2.6388  \\ 
b3    &  459194783490756864  &  35.39133923387  &  59.11965513235  &  51.2  &  15.578  &  $-$13.764  &  2.6504  \\ 
b4    &  458990239966637952  &  38.27197917057  &  58.81579898105  & 142.1  &  14.407  &  $-$14.858  &  2.6463  \\ 
b5    &  506840844266935936  &  33.97434127517  &  58.80312503169  &  28.7  &  15.721  &  $-$13.472  &  2.6606  \\ 
b6    &  459222236921401728  &  35.07063047584  &  59.44654655126  &  41.8  &  15.405  &  $-$14.625  &  2.7479  \\ 
b7    &  507255909903255552  &  34.84053716405  &  59.45844918367  &  35.3  &  15.780  &  $-$13.580  &  2.7361  \\ 
b8    &  459218938386541952  &  35.16801877868  &  59.38902174977  &  44.0  &  15.332  &  $-$13.384  &  2.6124  \\ 
      &                      &                  &                  &        &          &             &          \\
r1    &  507146783374181632  &  32.13145197747  &  59.96836576552  &  64.7  &  14.563  &  $-$13.005  &  2.6649  \\ 
r2    &  507365246896272896  &  35.06444839762  &  59.87623805527  &  54.2  &  15.155  &  $-$14.354  &  2.7007  \\ 
r3    &  507252993628942720  &  34.05459164780  &  59.64717386226  &  24.6  &  15.341  &  $-$14.737  &  2.5698  \\ 
r4    &  506860674140167040  &  33.76117606478  &  59.09963634588  &  10.0  &  15.939  &  $-$13.423  &  2.6602  \\ 
r5    &  507320132561058816  &  33.33222692914  &  59.90338605120  &  40.3  &  15.592  &  $-$13.631  &  2.6496  \\ 
r6    &  507292679128459648  &  33.44378542242  &  59.39912184483  &  12.3  &  16.482  &  $-$12.394  &  2.6718  \\ 
r7    &  507314566281391872  &  33.15504274024  &  59.84529344667  &  39.1  &  15.737  &  $-$14.351  &  2.7005  \\ 
r8    &  507310202594666368  &  33.29582016188  &  59.75533553578  &  32.4  &  16.054  &  $-$14.473  &  2.6604  \\ 
      &                      &                  &                  &        &          &             &          \\
u1    &  506860055662585216  &  33.90923584021  &  59.09157220960  &  11.6  &  16.294  &  $-$13.680  &  2.6885  \\ 
u2    &  507226674069510656  &  34.15492336642  &  59.29460623870  &  12.5  &  16.491  &  $-$11.897  &  2.6186  \\ 
u3    &  507222619614549120  &  34.08846836495  &  59.05499281394  &  16.4  &  15.477  &  $-$14.016  &  2.5316  \\ 
u4    &  507300478788942336  &  33.82852920703  &  59.65770807066  &  23.6  &  15.742  &  $-$13.342  &  2.6230  \\ 
u5    &  507296046382572544  &  33.43614454737  &  59.55110040542  &  19.6  &  15.850  &  $-$14.199  &  2.6847  \\ 
u6    &  507327451184397568  &  34.11443819490  &  59.90860513398  &  40.1  &  15.392  &  $-$13.980  &  2.7318  \\ 
      &                      &                  &                  &        &          &             &          \\
g1    &  458067680993514880  &  37.96122969562  &  57.53016943124  & 168.4  &  15.583  &  $-$15.372  &  2.8234  \\ 
g2    &  459199662573391104  &  35.31451183206  &  59.24791059241  &  48.0  &  15.336  &  $-$14.112  &  2.6734  \\ 
g3    &  506910564480154624  &  33.36994098675  &  59.19599825280  &  12.4  &  16.185  &  $-$13.613  &  2.7007  \\ 
g4    &  507507702367267584  &  32.79846557813  &  59.98091354856  &  51.7  &  15.907  &  $-$12.985  &  2.6480  \\ 
g5    &  459118882826608640  &  35.40906474613  &  58.78418769010  &  58.8  &  15.333  &  $-$14.185  &  2.6013  \\ 
g6    &  507214579443494144  &  31.49382702819  &  60.27836579355  &  91.2  &  16.730  &  $-$13.491  &  2.6515  \\ 
g7    &  465132764751065984  &  38.41778408960  &  60.29074433497  & 153.7  &  14.627  &  $-$14.373  &  2.6353  \\ 
g8    &  459112148318029056  &  35.44132561197  &  58.57339552018  &  66.9  &  16.175  &  $-$12.680  &  2.8040  \\ 
g9    &  507240967720664576  &  33.81963029733  &  59.33494934281  &   4.6  &  17.456  &  $-$13.180  &  2.6970  \\ 
g10   &  507520106232760320  &  32.66560224473  &  60.07874260771  &  58.8  &  16.211  &  $-$13.392  &  2.6178  \\ 
      &                      &                  &                  &        &          &             &          \\
to1   &  459223645670638080  &  34.96867177658  &  59.52776540690  &  40.4  &  15.741  &  $-$14.267  &  2.6547  \\ 
to2   &  459214196742707328  &  34.89578467837  &  59.30823843080  &  35.2  &  15.511  &  $-$13.763  &  2.6627  \\ 
to3   &  507833157812702464  &  30.38334220690  &  59.80312184848  & 107.3  &  16.747  &  $-$13.447  &  2.5536  \\ 
to4   &  507254264940286848  &  34.59497162300  &  59.37600203694  &  26.7  &  15.436  &  $-$13.684  &  2.6403  \\ 
to5   &  507289792902722560  &  33.50346335389  &  59.40154837538  &  11.1  &  16.080  &  $-$13.635  &  2.6827  \\ 
to6   &  458031294029981056  &  37.33987184386  &  57.08988530288  & 173.0  &  15.551  &  $-$14.335  &  2.7773  \\ 
to7   &  507289036996187392  &  33.67587890681  &  59.39470155470  &   8.0  &  16.136  &  $-$13.725  &  2.6887  \\ 
to8   &  459047616427062656  &  36.71980324154  &  59.11633877589  &  91.7  &  15.411  &  $-$14.421  &  2.6387  \\ 
to9   &  459349814627528832  &  35.90280215246  &  59.92113484362  &  76.3  &  15.000  &  $-$13.755  &  2.6427  \\ 
to10  &  507270860693863552  &  34.39585873306  &  59.53594422822  &  25.5  &  15.508  &  $-$13.984  &  2.6759  \\ 
to11  &  507242926225718656  &  33.88222527863  &  59.38491919641  &   8.2  &  15.570  &  $-$13.842  &  2.7249  \\ 
to12  &  458972682139718400  &  37.53662882225  &  58.64313669405  & 123.0  &  14.500  &  $-$13.693  &  2.5444  \\ 
to13  &  507233683456274176  &  34.36804763539  &  59.36682677450  &  19.9  &  15.643  &  $-$13.655  &  2.6301  \\ 
to14  &  507299104399421696  &  33.71937266885  &  59.64359860910  &  22.6  &  15.665  &  $-$13.787  &  2.6654  \\ 

\hline
 
\end{tabular}
\end{center}
\end{table*}

\begin{table*}
\caption{Photometry for the stars observed spectroscopically in this work.}
\label{tab_fotom}
\begin{center}
\begin{tabular}{lccccccc}  
\hline\hline
Star   & $V$ & $(B-V)$ &  $J$   &  $H$  & $K_{\textrm{S}}$ & $G$ & $(G_{\textrm{BP}}-G_{\textrm{RP}})$ \\ %& $E(G_{\textrm{BP}}-G_{\textrm{RP}})$   \\  
\hline
b1   &   9.968  &  0.290  &  9.239  &   9.079  &   9.028  &   9.789  &  0.429  \\ %&  0.486    \\ %
b2   &  10.067  &  0.287  &  9.278  &   9.191  &   9.158  &   9.976  &  0.484  \\ %&  0.586    \\
b3   &  10.443  &  0.381  &  9.673  &   9.542  &   9.493  &  10.438  &  0.562  \\ %&  0.566    \\
b4   &  10.760  &  0.432  &  9.801  &   9.653  &   9.594  &  10.681  &  0.638  \\ %&  0.551    \\
b5   &  11.223  &  0.493  & 10.081  &   9.891  &   9.871  &  11.061  &  0.731  \\ %&  0.604    \\
b6   &  12.459  &  0.684  & 11.132  &  10.795  &  10.705  &  12.308  &  0.927  \\ %&  0.203    \\  
b7   &  12.650  &  0.696  & 11.244  &  10.925  &  10.913  &  12.483  &  0.958  \\ %&  0.244    \\  % anyadir tambien extinciones
b8   &  13.204  &  0.781  & 11.676  &  11.344  &  11.210  &  13.000  &  1.049  \\ %&  0.267    \\
     &          &         &         &          &          &          &         \\
r1   &   9.522  &  0.340  &  8.679  &   8.576  &   8.530  &   9.410  &  0.502  \\ %&  0.353    \\
r2   &  10.019  &  0.355  &  8.992  &   8.919  &   8.820  &   9.825  &  0.602  \\ %&  0.379    \\
r3   &  10.271  &  0.492  &  9.215  &   9.102  &   9.022  &  10.165  &  0.705  \\ %&  0.511    \\
r4   &  10.835  &  0.563  &  9.622  &   9.401  &   9.358  &  10.691  &  0.800  \\ %&  0.440    \\
r5   &  11.280  &  0.650  & 10.108  &   9.873  &   9.808  &  11.172  &  0.835  \\ %&  0.659    \\
r6   &  11.793  &  0.675  & 10.414  &  10.127  &  10.045  &  11.602  &  0.937  \\ %&  0.504    \\
r7   &  12.610  &  0.855  & 11.124  &  10.739  &  10.657  &  12.405  &  1.075  \\ %&  0.402    \\
r8   &  13.075  &  0.891  & 11.481  &  11.079  &  11.035  &  12.854  &  1.122  \\ %&  0.391    \\ 
     &          &         &         &          &          &          &         \\
u1   &   9.559  &  0.564  &  8.359  &   8.277  &   8.158  &   9.379  &  0.770  \\ %&  0.588    \\
u2   &  10.085  &  0.569  &  8.752  &   8.553  &   8.460  &   9.849  &  0.844  \\ %&  0.701    \\
u3   &  10.725  &  0.680  &  9.315  &   9.119  &   9.017  &  10.529  &  0.952  \\ %&  0.556    \\
u4   &  11.570  &  0.867  &  9.889  &   9.689  &   9.573  &  11.334  &  1.132  \\ %&  0.811    \\
u5   &  12.078  &  0.923  & 10.277  &  10.000  &   9.855  &  11.784  &  1.198  \\ %&  0.613    \\
u6   &  13.111  &  1.000  & 11.074  &  10.776  &  10.652  &  12.769  &  1.334  \\ %&  0.545    \\
     &          &         &         &          &          &          &         \\
g1   &   7.132  &  1.348  &  4.764  &   3.920  &   3.730  &   6.689  &  1.476  \\ %&  0.099    \\
g2   &   7.492  &  1.233  &  5.167  &   4.597  &   4.449  &   7.006  &  1.454  \\ %&  0.227    \\
g3   &   7.633  &  1.189  &  5.394  &   4.889  &   4.718  &   7.199  &  1.433  \\ %&  0.149    \\
g4   &   8.222  &  1.234  &  5.915  &   5.367  &   5.213  &   7.782  &  1.466  \\ %&  0.140    \\
g5   &   8.201  &  1.085  &  6.153  &   5.657  &   5.520  &   7.819  &  1.306  \\ %&  0.113    \\
g6   &   8.402  &  1.410  &  5.878  &   5.286  &   5.148  &   7.879  &  1.582  \\ %&  0.217    \\
g7   &   8.401  &  1.192  &  6.117  &   5.593  &   5.438  &   7.960  &  1.441  \\ %&  0.129    \\
g8   &   8.359  &  1.003  &  6.410  &   5.916  &   5.813  &   7.999  &  1.252  \\ %&  0.254    \\
g9   &   8.892  &  1.342  &  6.195  &   5.632  &   5.450  &   8.309  &  1.701  \\ %&  0.359    \\
g10  &   8.975  &  1.144  &  6.639  &   6.131  &   5.986  &   8.497  &  1.458  \\ %&  0.269    \\
     &          &         &         &          &          &          &         \\
to1  &   8.220  &  0.215  &  7.606  &   7.669  &   7.585  &   8.134  &  0.317  \\ %&  0.313    \\
to2  &   8.291  &  0.297  &  7.517  &   7.476  &   7.411  &   8.194  &  0.447  \\ %&  0.687    \\
to3  &   8.324  &  0.236  &  7.662  &   7.650  &   7.571  &   8.241  &  0.369  \\ %&  0.282    \\
to4  &   8.527  &  0.342  &  7.666  &   7.568  &   7.534  &   8.399  &  0.501  \\ %&  0.445    \\
to5  &   8.585  &  0.407  &  7.655  &   7.557  &   7.518  &   8.475  &  0.580  \\ %&  0.724    \\
to6  &   8.632  &  0.182  &  8.171  &   8.143  &   8.094  &   8.551  &  0.248  \\ %&  0.386    \\
to7  &   8.704  &  0.445  &  7.683  &   7.613  &   7.552  &   8.575  &  0.619  \\ %&  0.853    \\
to8  &   8.666  &  0.222  &  8.042  &   8.031  &   7.991  &   8.583  &  0.340  \\ %&  0.246    \\
to9  &   8.734  &  0.437  &  7.727  &   7.591  &   7.509  &   8.585  &  0.625  \\ %&  0.774    \\
to10 &   8.708  &  0.357  &  7.859  &   7.783  &   7.729  &   8.605  &  0.514  \\ %&  0.540    \\
to11 &   8.779  &  0.429  &  7.747  &   7.701  &   7.613  &   8.633  &  0.620  \\ %&  0.730    \\
to12 &   8.744  &  0.219  &  8.121  &   8.125  &   8.045  &   8.656  &  0.340  \\ %&  0.426    \\
to13 &   8.812  &  0.347  &  7.886  &   7.798  &   7.750  &   8.679  &  0.537  \\ %&  0.680    \\
to14 &   9.041  &  0.477  &  7.906  &   7.815  &   7.698  &   8.883  &  0.719  \\ %&  0.676    \\

\hline
 
\end{tabular}
\end{center}
%\begin{list}{}{}
%\item[]$^{*}$ Values estimated in this work.
%  \end{list}
\end{table*}

\begin{landscape}

\begin{table}
\caption{Chemical abundances, expressed as $A(X)$=log[$n(X)/n(H)$]+12, for MS stars in Stock\,2.}
\begin{center}
\begin{tabular}{lcccccccc} 
\hline\hline
X  &   b1   &  b2   &  b4   &  b5    &  b6    &  b7   &  b8  &  r3   \\       
\hline
C  &  8.66 $\pm$ 0.09  &  8.67 $\pm$ 0.18  &  8.47 $\pm$ 0.08  &  8.65 $\pm$ 0.17  &         $\dots$       &  8.32 $\pm$ 0.11  &        $\dots$        &  8.35 $\pm$ 0.15  \\
O  &  8.57 $\pm$ 0.13  &  8.53 $\pm$ 0.11  &  8.41 $\pm$ 0.16  &         $\dots$       &         $\dots$       &        $\dots$        &        $\dots$        &  8.29 $\pm$ 0.15  \\
Na &         $\dots$       &         $\dots$       &  6.42 $\pm$ 0.15  &  6.44 $\pm$ 0.15  &  6.29 $\pm$ 0.15  &  6.23 $\pm$ 0.09  &  6.26 $\pm$ 0.04  &  6.15 $\pm$ 0.13  \\
Mg &  7.54 $\pm$ 0.07  &  7.55 $\pm$ 0.15  &  7.52 $\pm$ 0.13  &  7.60 $\pm$ 0.15  &  7.33 $\pm$ 0.05  &  7.40 $\pm$ 0.09  &  7.47 $\pm$ 0.13  &  7.53 $\pm$ 0.12  \\
Al &         $\dots$       &         $\dots$       &         $\dots$       &         $\dots$       &         $\dots$       &  6.10 $\pm$ 0.15  &  6.30 $\pm$ 0.10  &         $\dots$       \\
Si &  7.56 $\pm$ 0.19  &  7.66 $\pm$ 0.12  &  7.57 $\pm$ 0.10  &         $\dots$       &  7.46 $\pm$ 0.15  &  7.44 $\pm$ 0.09  &  7.58 $\pm$ 0.05  &  7.59 $\pm$ 0.11  \\
S  &  7.29 $\pm$ 0.15  &         $\dots$       &  7.23 $\pm$ 0.15  &         $\dots$       &         $\dots$       &  7.41 $\pm$ 0.15  &        $\dots$        &  7.45 $\pm$ 0.15  \\
Ca &  6.33 $\pm$ 0.16  &  6.12 $\pm$ 0.14  &  6.19 $\pm$ 0.11  &  6.37 $\pm$ 0.10  &  6.48 $\pm$ 0.12  &  6.31 $\pm$ 0.15  &  6.35 $\pm$ 0.12  &        $\dots$        \\
Sc &  3.38 $\pm$ 0.10  &  3.19 $\pm$ 0.15  &  3.24 $\pm$ 0.15  &         $\dots$       &  3.01 $\pm$ 0.15  &  2.99 $\pm$ 0.13  &  3.18 $\pm$ 0.07  &  3.19 $\pm$ 0.15  \\
Ti &  4.91 $\pm$ 0.17  &  4.85 $\pm$ 0.19  &  5.00 $\pm$ 0.17  &         $\dots$       &  4.99 $\pm$ 0.15  &  4.76 $\pm$ 0.06  &  4.89 $\pm$ 0.11  &  4.62 $\pm$ 0.10  \\
V  &  4.21 $\pm$ 0.15  &  4.09 $\pm$ 0.14  &  4.13 $\pm$ 0.15  &         $\dots$       &  3.96 $\pm$ 0.15  &  3.81 $\pm$ 0.19  &  4.08 $\pm$ 0.18  &  4.41 $\pm$ 0.15  \\
Cr &  5.54 $\pm$ 0.11  &  5.73 $\pm$ 0.14  &  5.56 $\pm$ 0.12  &         $\dots$       &  5.53 $\pm$ 0.14  &  5.50 $\pm$ 0.11  &  5.68 $\pm$ 0.15  &  5.59 $\pm$ 0.14  \\
Mn &  5.41 $\pm$ 0.15  &  5.43 $\pm$ 0.14  &  5.35 $\pm$ 0.15  &  5.35 $\pm$ 0.15  &  5.20 $\pm$ 0.16  &  5.17 $\pm$ 0.15  &  5.42 $\pm$ 0.20  &  5.35 $\pm$ 0.15  \\
Fe &  7.49 $\pm$ 0.19  &  7.61 $\pm$ 0.12  &  7.02 $\pm$ 0.15  &  7.40 $\pm$ 0.10  &  7.39 $\pm$ 0.15  &  7.31 $\pm$ 0.10  &  7.25 $\pm$ 0.12  &  7.06 $\pm$ 0.18  \\
Co &  4.88 $\pm$ 0.15  &         $\dots$       &  4.88 $\pm$ 0.15  &  4.88 $\pm$ 0.10  &  5.00 $\pm$ 0.15  &  4.71 $\pm$ 0.15  &  4.89 $\pm$ 0.18  &  5.07 $\pm$ 0.09  \\
Ni &  6.28 $\pm$ 0.14  &  6.50 $\pm$ 0.09  &  6.07 $\pm$ 0.17  &         $\dots$       &  6.12 $\pm$ 0.14  &  6.02 $\pm$ 0.17  &  6.07 $\pm$ 0.11  &  6.25 $\pm$ 0.20  \\
Cu &         $\dots$       &         $\dots$       &         $\dots$       &         $\dots$       &         $\dots$       &  4.07 $\pm$ 0.11  &  4.00 $\pm$ 0.16  &        $\dots$        \\
Zn &         $\dots$       &         $\dots$       &  4.31 $\pm$ 0.15  &         $\dots$       &  4.32 $\pm$ 0.15  &  4.32 $\pm$ 0.06  &        $\dots$        &  4.58 $\pm$ 0.15  \\
Sr &  2.86 $\pm$ 0.15  &  3.05 $\pm$ 0.13  &  3.01 $\pm$ 0.15  &         $\dots$       &  2.90 $\pm$ 0.15  &  3.02 $\pm$ 0.13  &  2.90 $\pm$ 0.10  &  2.90 $\pm$ 0.04  \\
Y  &  2.26 $\pm$ 0.19  &  2.57 $\pm$ 0.19  &  2.30 $\pm$ 0.11  &         $\dots$       &  2.14 $\pm$ 0.08  &  2.21 $\pm$ 0.19  &  2 .19 $\pm$ 0.09 &  2.25 $\pm$ 0.11  \\
Zr &         $\dots$       &         $\dots$       &  2.56 $\pm$ 0.15  &         $\dots$       &  2.56 $\pm$ 0.15  &  2.64 $\pm$ 0.11  &  2.76 $\pm$ 0.15  &  2.40 $\pm$ 0.16  \\
Ba &  1.84 $\pm$ 0.15  &  2.64 $\pm$ 0.12  &  1.63 $\pm$ 0.19  &         $\dots$       &  1.97 $\pm$ 0.15  &  2.03 $\pm$ 0.19  &  2.09 $\pm$ 0.15  &  1.95 $\pm$ 0.11  \\
   &                   &                   &                   &                   &                   &                   &                   &                   \\
\hline\hline
X  &  r5   &  r6  & r7  & r8  &  u3  & u4  & u5  & u6 \\       
\hline
C  &  8.66 $\pm$ 0.11  &  8.50 $\pm$ 0.13  &  8.61 $\pm$ 0.15  &  8.27 $\pm$ 0.08  &  8.41 $\pm$ 0.15  &  8.42 $\pm$ 0.11  &  8.07 $\pm$ 0.15  &  8.62 $\pm$ 0.17  \\
O  &  8.45 $\pm$ 0.08  &  8.79 $\pm$ 0.18  &         $\dots$       &         $\dots$       &  8.65 $\pm$ 0.04  &  8.78 $\pm$ 0.11  &          $\dots$      &         $\dots$       \\
Na &         $\dots$       &  6.44 $\pm$ 0.15  &  6.11 $\pm$ 0.14  &  6.13 $\pm$ 0.08  &  6.21 $\pm$ 0.14  &        $\dots$        &  6.29 $\pm$ 0.09  &  6.20 $\pm$ 0.17  \\
Mg &  7.44 $\pm$ 0.13  &  7.40 $\pm$ 0.10  &  7.40 $\pm$ 0.13  &  7.43 $\pm$ 0.11  &  7.32 $\pm$ 0.11  &  7.59 $\pm$ 0.06  &  7.56 $\pm$ 0.15  &  7.42 $\pm$ 0.10  \\
Al &         $\dots$       &  6.45 $\pm$ 0.10  &        $\dots$        &  6.50 $\pm$ 0.10  &        $\dots$        &        $\dots$        &         $\dots$       &  6.30 $\pm$ 0.15  \\
Si &  7.58 $\pm$ 0.18  &  7.48 $\pm$ 0.12  &  7.43 $\pm$ 0.15  &  7.51 $\pm$ 0.09  &  7.33 $\pm$ 0.12  &  7.61 $\pm$ 0.15  &  7.61 $\pm$ 0.19  &  7.45 $\pm$ 0.17  \\
S  &         $\dots$       &  7.38 $\pm$ 0.12  &  7.33 $\pm$ 0.19  &  7.19 $\pm$ 0.20  &  7.35 $\pm$ 0.08  &  7.29 $\pm$ 0.15  &         $\dots$       &  7.20 $\pm$ 0.14  \\
Ca &  6.35 $\pm$ 0.19  &  6.20 $\pm$ 0.09  &  6.33 $\pm$ 0.11  &  6.42 $\pm$ 0.10  &  6.12 $\pm$ 0.17  &  6.27 $\pm$ 0.12  &  6.37 $\pm$ 0.18  &  6.31 $\pm$ 0.11  \\
Sc &  3.13 $\pm$ 0.10  &  2.93 $\pm$ 0.10  &  3.23 $\pm$ 0.17  &  3.06 $\pm$ 0.12  &  3.19 $\pm$ 0.09  &  3.07 $\pm$ 0.11  &  3.13 $\pm$ 0.10  &  3.18 $\pm$ 0.15  \\
Ti &  4.75 $\pm$ 0.05  &  4.82 $\pm$ 0.14  &  4.73 $\pm$ 0.15  &  4.91 $\pm$ 0.12  &  4.56 $\pm$ 0.14  &  4.73 $\pm$ 0.10  &  4.99 $\pm$ 0.13  &  4.61 $\pm$ 0.14  \\
V  &  4.04 $\pm$ 0.14  &  3.98 $\pm$ 0.16  &  4.23 $\pm$ 0.11  &  4.01 $\pm$ 0.16  &  4.05 $\pm$ 0.08  &  4.17 $\pm$ 0.19  &  4.22 $\pm$ 0.15  &  4.06 $\pm$ 0.16  \\
Cr &  5.52 $\pm$ 0.16  &  5.48 $\pm$ 0.11  &  5.34 $\pm$ 0.12  &  5.61 $\pm$ 0.09  &  5.29 $\pm$ 0.11  &  5.64 $\pm$ 0.09  &  5.63 $\pm$ 0.08  &  5.28 $\pm$ 0.11  \\
Mn &  4.97 $\pm$ 0.13  &  5.13 $\pm$ 0.14  &  5.27 $\pm$ 0.11  &  5.39 $\pm$ 0.17  &  5.05 $\pm$ 0.13  &  5.10 $\pm$ 0.15  &  5.48 $\pm$ 0.14  &  5.22 $\pm$ 0.15  \\
Fe &  7.20 $\pm$ 0.10  &  7.15 $\pm$ 0.14  &  7.14 $\pm$ 0.12  &  7.43 $\pm$ 0.08  &  7.10 $\pm$ 0.15  &  7.37 $\pm$ 0.16  &  7.33 $\pm$ 0.13  &  7.18 $\pm$ 0.15  \\
Co &  5.23 $\pm$ 0.15  &  5.10 $\pm$ 0.20  &  5.16 $\pm$ 0.09  &  4.89 $\pm$ 0.18  &  5.00 $\pm$ 0.20  &  5.03 $\pm$ 0.15  &  5.32 $\pm$ 0.11  &  4.91 $\pm$ 0.08  \\
Ni &  6.08 $\pm$ 0.15  &  5.90 $\pm$ 0.11  &  5.91 $\pm$ 0.17  &  6.12 $\pm$ 0.18  &  6.00 $\pm$ 0.11  &  6.25 $\pm$ 0.17  &  6.21 $\pm$ 0.08  &  5.97 $\pm$ 0.17  \\
Cu &  4.17 $\pm$ 0.15  &  4.30 $\pm$ 0.15  &  4.11 $\pm$ 0.15  &  4.08 $\pm$ 0.04  &        $\dots$        &        $\dots$        &         $\dots$       &  4.24 $\pm$ 0.15  \\
Zn &        $\dots$        &  4.22 $\pm$ 0.11  &  4.15 $\pm$ 0.13  &  4.43 $\pm$ 0.10  &        $\dots$        &        $\dots$        &         $\dots$       &  4.35 $\pm$ 0.15  \\
Sr &        $\dots$        &  2.93 $\pm$ 0.10  &         $\dots$       &  3.00 $\pm$ 0.04  &  2.89 $\pm$ 0.04  &  2.87 $\pm$ 0.15  &         $\dots$       &  3.03 $\pm$ 0.15  \\
Y  &        $\dots$        &  2.29 $\pm$ 0.11  &  2.50 $\pm$ 0.05  &  2.28 $\pm$ 0.19  &  2.48 $\pm$ 0.04  &  2.24 $\pm$ 0.19  &  2.20 $\pm$ 0.15  &  2.00 $\pm$ 0.15  \\
Zr &  2.56 $\pm$ 0.14  &  2.66 $\pm$ 0.14  &  2.62 $\pm$ 0.17  &  2.70 $\pm$ 0.19  &  2.82 $\pm$ 0.08  &        $\dots$        &         $\dots$       &  2.57 $\pm$ 0.14  \\
Ba &  1.61 $\pm$ 0.16  &  1.89 $\pm$ 0.14  &  1.91 $\pm$ 0.08  &  2.03 $\pm$ 0.08  &  1.39 $\pm$ 0.14  &  2.06 $\pm$ 0.04  &  2.14 $\pm$ 0.18  &  2.05 $\pm$ 0.13  \\

\hline
\end{tabular}

\label{abund_ms}
\end{center}
\end{table}

\end{landscape}

\begin{landscape}

\begin{table}
\caption{Chemical abundances, expressed as $A(X)$=log[$n(X)/n(H)$]+12, for giants in Stock\,2.}
\begin{center}
\begin{tabular}{lcccccccccc} 
\hline\hline
X   &      g1   &   g2    &    g3    &   g4    &     g5    &   g6    &   g7   &    g8   &   g9   &   g10        \\       
\hline
%C   &        $\dots$        &         $\dots$       &  8.80 $\pm$ 0.15  &  9.38 $\pm$ 0.10  &  9.23 $\pm$ 0.10  &  8.73 $\pm$ 0.10  &  8.98 $\pm$ 0.10  &  9.23 $\pm$ 0.10  &        $\dots$       &         $\dots$       \\
%O   &  8.84 $\pm$ 0.06  &  8.85 $\pm$ 0.06  &  8.94 $\pm$ 0.15  &  9.00 $\pm$ 0.08  &  8.95 $\pm$ 0.05  &  9.07 $\pm$ 0.15  &  8.95 $\pm$ 0.10  &  8.84 $\pm$ 0.05  & 8.91 $\pm$ 0.10  &  8.87 $\pm$ 0.07  \\
Na  &  6.35 $\pm$ 0.08  &  6.38 $\pm$ 0.15  &  6.56 $\pm$ 0.13  &  6.49 $\pm$ 0.13  &  6.39 $\pm$ 0.10  &  6.56 $\pm$ 0.15  &  6.50 $\pm$ 0.10  &  6.31 $\pm$ 0.10  & 6.46 $\pm$ 0.13  &  6.35 $\pm$ 0.06  \\
Mg  &  7.59 $\pm$ 0.12  &  7.52 $\pm$ 0.13  &  7.58 $\pm$ 0.09  &  7.55 $\pm$ 0.03  &  7.51 $\pm$ 0.12  &  7.54 $\pm$ 0.06  &  7.63 $\pm$ 0.15  &  7.47 $\pm$ 0.15  & 7.71 $\pm$ 0.18  &  7.52 $\pm$ 0.16  \\
Al  &  6.37 $\pm$ 0.08  &  6.46 $\pm$ 0.09  &  6.41 $\pm$ 0.02  &  6.48 $\pm$ 0.17  &  6.56 $\pm$ 0.13  &  6.41 $\pm$ 0.12  &  6.48 $\pm$ 0.14  &  6.56 $\pm$ 0.13  & 6.68 $\pm$ 0.10  &  6.50 $\pm$ 0.10  \\
Si  &  7.63 $\pm$ 0.09  &  7.58 $\pm$ 0.12  &  7.60 $\pm$ 0.12  &  7.57 $\pm$ 0.13  &  7.54 $\pm$ 0.09  &  7.58 $\pm$ 0.11  &  7.57 $\pm$ 0.12  &  7.48 $\pm$ 0.12  & 7.57 $\pm$ 0.11  &  7.54 $\pm$ 0.13  \\
S   &        $\dots$        &         $\dots$       &  7.49 $\pm$ 0.13  &  7.41 $\pm$ 0.18  &  7.54 $\pm$ 0.15  &  7.47 $\pm$ 0.15  &  7.47 $\pm$ 0.15  &  7.47 $\pm$ 0.16  & 7.47 $\pm$ 0.26  &  7.44 $\pm$ 0.12  \\
Ca  &  6.24 $\pm$ 0.15  &  6.20 $\pm$ 0.15  &  6.28 $\pm$ 0.15  &  6.30 $\pm$ 0.15  &  6.30 $\pm$ 0.14  &  6.25 $\pm$ 0.10  &  6.21 $\pm$ 0.12  &  6.22 $\pm$ 0.15  & 6.20 $\pm$ 0.15  &  6.11 $\pm$ 0.15  \\
Sc  &  3.09 $\pm$ 0.07  &  3.18 $\pm$ 0.14  &  3.21 $\pm$ 0.12  &  3.24 $\pm$ 0.12  &  3.16 $\pm$ 0.10  &  3.16 $\pm$ 0.13  &  3.23 $\pm$ 0.12  &  3.23 $\pm$ 0.14  & 3.23 $\pm$ 0.14  &  3.16 $\pm$ 0.13  \\
Ti  &  4.78 $\pm$ 0.15  &  4.77 $\pm$ 0.15  &  4.94 $\pm$ 0.15  &  4.90 $\pm$ 0.10  &  4.86 $\pm$ 0.13  &  4.94 $\pm$ 0.15  &  4.87 $\pm$ 0.11  &  4.79 $\pm$ 0.15  & 4.84 $\pm$ 0.14  &  4.79 $\pm$ 0.15  \\
V   &  3.90 $\pm$ 0.16  &  3.93 $\pm$ 0.09  &  3.97 $\pm$ 0.15  &  3.94 $\pm$ 0.15  &  3.90 $\pm$ 0.13  &  3.90 $\pm$ 0.15  &  3.97 $\pm$ 0.15  &  3.88 $\pm$ 0.12  & 3.96 $\pm$ 0.08  &  3.88 $\pm$ 0.12  \\
Cr  &  5.64 $\pm$ 0.12  &  5.65 $\pm$ 0.13  &  5.70 $\pm$ 0.12  &  5.73 $\pm$ 0.12  &  5.62 $\pm$ 0.07  &  5.72 $\pm$ 0.12  &  5.73 $\pm$ 0.12  &  5.63 $\pm$ 0.05  & 5.68 $\pm$ 0.12  &  5.65 $\pm$ 0.14  \\
Mn  &  5.29 $\pm$ 0.10  &  5.30 $\pm$ 0.12  &  5.34 $\pm$ 0.11  &  5.29 $\pm$ 0.10  &  5.34 $\pm$ 0.08  &  5.31 $\pm$ 0.09  &  5.28 $\pm$ 0.11  &  5.28 $\pm$ 0.13  & 5.27 $\pm$ 0.11  &  5.28 $\pm$ 0.11  \\
Fe  &  7.30 $\pm$ 0.19  &  7.30 $\pm$ 0.13  &  7.45 $\pm$ 0.09  &  7.35 $\pm$ 0.14  &  7.29 $\pm$ 0.11  &  7.40 $\pm$ 0.13  &  7.31 $\pm$ 0.12  &  7.34 $\pm$ 0.12  & 7.28 $\pm$ 0.15  &  7.30 $\pm$ 0.14  \\
Co  &  4.83 $\pm$ 0.13  &  4.87 $\pm$ 0.07  &  4.81 $\pm$ 0.10  &  4.83 $\pm$ 0.14  &  4.87 $\pm$ 0.13  &  4.82 $\pm$ 0.13  &  4.82 $\pm$ 0.11  &  4.85 $\pm$ 0.13  & 4.93 $\pm$ 0.13  &  4.78 $\pm$ 0.12  \\
Ni  &  6.04 $\pm$ 0.11  &  6.00 $\pm$ 0.15  &  6.17 $\pm$ 0.13  &  6.17 $\pm$ 0.14  &  6.09 $\pm$ 0.14  &  6.19 $\pm$ 0.14  &  6.20 $\pm$ 0.13  &  6.03 $\pm$ 0.13  & 6.09 $\pm$ 0.15  &  6.00 $\pm$ 0.09  \\
Cu  &  3.88 $\pm$ 0.15  &  3.96 $\pm$ 0.14  &  4.11 $\pm$ 0.15  &  3.99 $\pm$ 0.15  &  3.96 $\pm$ 0.11  &  3.96 $\pm$ 0.15  &  3.99 $\pm$ 0.12  &  3.90 $\pm$ 0.11  & 3.93 $\pm$ 0.15  &  3.87 $\pm$ 0.15  \\
Zn  &  4.53 $\pm$ 0.15  &  4.25 $\pm$ 0.09  &  4.42 $\pm$ 0.12  &  4.35 $\pm$ 0.08  &  4.32 $\pm$ 0.10  &  4.47 $\pm$ 0.10  &  4.55 $\pm$ 0.12  &  4.49 $\pm$ 0.09  & 4.39 $\pm$ 0.10  &  4.33 $\pm$ 0.15  \\
Sr  &  3.03 $\pm$ 0.20  &  3.31 $\pm$ 0.13  &  3.53 $\pm$ 0.13  &  3.43 $\pm$ 0.10  &  3.51 $\pm$ 0.10  &  3.31 $\pm$ 0.14  &  3.36 $\pm$ 0.14  &  3.46 $\pm$ 0.11  & 3.30 $\pm$ 0.15  &  3.43 $\pm$ 0.10  \\
Y   &  2.37 $\pm$ 0.13  &  2.40 $\pm$ 0.12  &  2.36 $\pm$ 0.13  &  2.30 $\pm$ 0.13  &  2.36 $\pm$ 0.12  &  2.23 $\pm$ 0.13  &  2.13 $\pm$ 0.15  &  2.34 $\pm$ 0.14  & 2.21 $\pm$ 0.13  &  2.39 $\pm$ 0.18  \\
Zr  &  2.59 $\pm$ 0.15  &  2.71 $\pm$ 0.05  &  2.62 $\pm$ 0.08  &  2.64 $\pm$ 0.12  &  2.66 $\pm$ 0.12  &  2.59 $\pm$ 0.05  &  2.64 $\pm$ 0.14  &  2.62 $\pm$ 0.14  & 2.66 $\pm$ 0.12  &  2.64 $\pm$ 0.14  \\
Ba  &  2.47 $\pm$ 0.18  &  2.42 $\pm$ 0.05  &  2.34 $\pm$ 0.06  &  2.30 $\pm$ 0.04  &  2.25 $\pm$ 0.09  &  2.47 $\pm$ 0.14  &  2.47 $\pm$ 0.14  &  2.47 $\pm$ 0.10  & 2.40 $\pm$ 0.18  &  2.42 $\pm$ 0.12  \\
\hline                     

\end{tabular}

\label{abund_gig}
\end{center}
\end{table}

\end{landscape}

\end{document}